\documentstyle[12pt,dina4,amsfont12]{article}  
\input psfig.sty  
\begin{document}  
\newcommand{\pl}[2]{\frac{\partial#1}{\partial#2}}  
\newcommand{\Ta}{{\cal A}}  
\newcommand{\Tb}{{\cal B}}  
\newcommand{\Te}{{\cal E}}  
\newcommand{\bu}{{\bf u} } 
\newcommand{\p}{\partial}  
\newcommand{\og}{\omega}  
\newcommand{\Og}{\Omega}  
\newcommand{\fl}[2]{\frac{#1}{#2}}  
\newcommand{\dt}{\delta}  
\newcommand{\tm}{\times}  
\newcommand{\sm}{\setminus}  
\newcommand{\nn}{\nonumber}  
\newcommand{\ap}{\alpha}  
\newcommand{\bt}{\beta}  
\newcommand{\ld}{\lambda}  
\newcommand{\Gm}{\Gamma}  
\newcommand{\gm}{\gamma}  
\newcommand{\vp}{\varphi}  
\newcommand{\tht}{\theta}  
\newcommand{\ift}{\infty}  
\newcommand{\vep}{\varepsilon}  
\newcommand{\ep}{\epsilon}  
\newcommand{\kp}{\kappa}  
\newcommand{\Dt}{\Delta}  
\newcommand{\Sg}{\Sigma}  
\newcommand{\fa}{\forall}  
\newcommand{\sg}{\sigma}  
\newcommand{\ept}{\emptyset}  
\newcommand{\btd}{\nabla}   
\newcommand{\btu}{\Delta}  
\newcommand{\tg}{\triangle}  
\newcommand{\Th}{{\cal T}_h}  
\newcommand{\ged}{\qquad \Box}  
\newcommand{\bgv}{\Bigg\vert}  
\renewcommand{\theequation}{\arabic{section}.\arabic{equation}}  
\newcommand{\be}{\begin{equation}}  
\newcommand{\ee}{\end{equation}}  
\newcommand{\ba}{\begin{array}}  
\newcommand{\ea}{\end{array}}  
\newcommand{\bea}{\begin{eqnarray}}  
\newcommand{\eea}{\end{eqnarray}}  
\newcommand{\beas}{\begin{eqnarray*}}  
\newcommand{\eeas}{\end{eqnarray*}}  
\newcommand{\dpm}{\displaystyle}  
\newtheorem{theorem}{Theorem}[section]  
\newtheorem{lemma}{Lemma}[section]  
\newtheorem{remark}{Remark}[section]  
\newcommand{\Gmu}{\Gm_{_U}}  
\newcommand{\Gml}{\Gm_{_L}}  
\newcommand{\Gme}{\Gm_e}  
\newcommand{\Gmi}{\Gm_i}  
\newcommand{\lN}{{_N}}  
\newcommand{\tld}[1]{\~{#1}}  
\newcommand{\td}[1]{\tilde{#1}}  
\newcommand{\um}{\mu}  
\newcommand{\bx}{{\bf x} } 
\newcommand{\Eb}{E_\kp }
 \newcommand{\mub}{\mu_\kp }

\title{Ground state solution of 
Bose-Einstein condensate by directly minimizing the energy
 functional }
\author{ {\it Weizhu Bao}  
\thanks{Email address:  bao@cz3.nus.edu.sg. Fax: 65-67746756}\\  
Department of Computational Science\\ 
National University of Singapore, Singapore 117543.\\  
\\
{\it Weijun Tang }
\thanks{Email address: tang\_tangwj@yahoo.com.}\\
Department of Computational Science, \\
National University of Singapore, Singapore 117543\\
and Institute of Applied Physics and Computational Mathematics\\
P.O. Box 8009, Beijing, 100088, P.R. China.\\ 
}  
  
\date{}  
\maketitle  
  
\begin{abstract}  
  In this paper, we propose a new numerical method to compute
the ground state solution of trapped interacting
Bose-Einstein condensation (BEC)
at zero or very low temperature by directly minimizing the energy
 functional via finite element approximation.  
As preparatory steps we begin with the 3d Gross-Pitaevskii equation (GPE),
scale it to get a three-parameter model and show how to reduce
it to 2d and 1d GPEs.
The ground state solution is formulated by 
 minimizing the energy  functional
under a constraint,  which is discretized by the finite
element method. The finite element approximation for 1d, 2d with radial
symmetry and 3d with spherical symmetry and cylindrical symmetry
 are presented in detail and 
approximate ground state solutions, which are used as
initial guess in our practical numerical computation of 
the minimization problem, of the GPE in two extreme
regimes: very weak interactions and strong repulsive interactions 
are provided.  
Numerical results in 1d, 2d with radial symmetry and 3d with spherical
symmetry and cylindrical symmetry 
for atoms ranging up to millions in the condensation
are reported to demonstrate the novel numerical method.
 Furthermore, comparisons between the ground state solutions
and their Thomas-Fermi approximations are also reported.
Extension of the numerical method to compute the
excited states of GPE is also presented. 
\end{abstract}  

  {\sl Key Words:} Bose-Einstein condensation (BEC), ground state solution,
Gross-Pitaevskii equation (GPE), energy functional, 
finite element approximation, Thomas-Fermi approximation.

\section{Introduction}\label{si}  
\setcounter{equation}{0}  
    
  Recently, there have been experiments  of 
Bose-Einstein condensation (BEC) in dilute bosonic atoms
(alkali and hydrogen atoms) employing magnetic traps at 
ultra-low temperatures \cite{Anderson,Bradley,Ensher}. 
 The condensation 
can consist of few thousand to millions of atoms confined by the trap 
potential. This peculiar state 
of matter, whose existence was postulated  back in the 1920s by Bose 
\cite{Bose} and Einstein \cite{Einstein}, 
exhibits several characteristics which set 
it apart from other 
homogeneous condensed matter systems \cite{LL,Cornell}. 
In fact, besides internal
interactions, the macroscopic behavior of BEC matter is highly 
sensitive to the shape of the 
external trapping potential. Theoretical predictions 
of the properties of BEC matter can now be compared with experimental
data by adjusting some tunable external parameters, such as the trap
frequency and/or aspect ratio. Needless to say, these dramatic progresses
on the experimental front are stimulation a corresponding 
wave of activity on both the theoretical and the numerical fronts. 

   The properties of a BEC at temperatures $T$ very much smaller
than the critical temperature $T_c$ \cite{LL,Griffin} are usually 
described by the nonlinear Schr\"{o}dinger equation (NLSE) for
the macroscopic wave function known as the 
Gross-Pitaevskii equation (GPE) \cite{Gross,Pit}. Note that 
equations very similar to the GPE also appear in nonlinear optics
where an index of refraction which depends on the light intensity 
leads to a nonlinear term like the one encountered in the GPE.

   There has been a series of recent studies which deal
with the numerical solution of the time-independent GPE for ground state
and the time-dependent GPE for finding the dynamics of a BEC. 
For numerical solution of time-dependent GPE, Bao et al. 
\cite{Bao,Bao1,Bao2}
presented a time-splitting spectral method,  Ruprecht et al.  \cite{Rup}
and Adhikari et al. \cite{Adh2,Adh3}
used the Crank-Nicolson finite difference method to compute the ground 
state solution and dynamics of GPE, Cerimele et. al. 
\cite{Cerim} proposed  a particle-inspired scheme. 
For ground state solution 
of GPE, Edwards et al. \cite{Edwards} presented a Runge-Kutta type method 
and used it to solve 1d and 3d with spherical symmetry time-independent GPE.
 Adhikari \cite{Adh,Adh1} 
used this approach to get the ground state solution
of GPE in 2d with radial symmetry. Other approaches include 
an explicit imaginary-time 
algorithm used  by Cerimele et al. \cite{Tosi2} and Chiofalo et al.
\cite{Tosi}, a direct
inversion in the iterated subspace (DIIS) used by Schneider et al. 
\cite{Feder}, and a simple analytical type method 
proposed by Dodd \cite{Dodd}.

  In this paper,  we propose a new numerical method to compute
the ground state solution of Bose-Einstein condensation 
by directly minimizing the energy functional through the finite element
discretization. 
We begin with the 3d Gross-Pitaevskii equation,
make it dimensionless to obtain a three-parameter model, 
show how to approximately reduce
it to a 2d GPE and a 1d GPE in certain limits.
The ground state solution is formulated by directly minimizing 
the energy functional under a constraint. Furthermore we present in detail 
the  finite element approximation  of the minimization problem for 1d, 
2d with radial symmetry and 3d with spherical symmetry and cylindrical 
symmetry (this is the most popular case in the current
experiments of BEC), and provide 
approximate ground state solutions,  which are used as
initial guess in our practical numerical computation, in two extreme
regimes: very weak interactions and strong repulsive interactions.
Numerical results in 1d, 2d with radial symmetry and 3d 
with spherical symmetry and cylindrical symmetry
 for atoms ranging up to millions in the condensation
are reported to demonstrate 
 the  numerical method. 
Furthermore, comparisons between the ground state solutions
and their Thomas-Fermi approximations are also reported by 
using our numerical method. Our numerical results
show that the Thomas-Fermi approximation is a `good'
approximation to the ground state solution in the strong
repulsive interaction regime, but a `worse' approximation
in the medium interaction regime. Convergence rate of the 
 Thomas-Fermi approximation to the ground state solution 
as a function of the number of atoms in the condensation
is also reported by our method. Furthermore, we
also extend our method to compute the excited states of GPE 
in 1d.

   The paper is organized as follows. In Section \ref{sgpel} we 
begin with the 3d GPE, scale it to get a three-parameter model,
show how to reduce it to lower dimensions.
In Section \ref{sna} we present a new method to compute
the ground state solution of a BEC
by directly minimizing the energy functional and provide the 
approximate ground state solution in two extreme
regimes: very weak interactions and strong repulsive interactions.
In Section \ref{s1d} we present detailed numerical formula and its 
finite element approximation for the ground state solution
of GPE in 1d, 2d with radial symmetry and 3d with spherical symmetry
(in these cases, it is reduced to a 1d problem),
and in section \ref{s2d} for 3d with cylindrical symmetry
(in this case, it is reduced to a 2d problem).
 In Section \ref{sne} we report on numerical results of the ground state
solution of BEC in 1d, 2d with radial symmetry and 3d with spherical
symmetry as well as cylindrical symmetry. 
In Section \ref{sc} some conclusions are drawn.

\section{Gross-Pitaevskii equation}\label{sgpel} 
\setcounter{equation}{0}  
 
  In this section, we will present the Gross-Pitaevskii equation
in three dimension, how to scale it to a 
three-parameter model and reduction to lower dimension.

   At zero or very low temperature, a BEC is well described by
the macroscopic wave function
$\psi(\bx,t)$ whose evolution is governed by a self-consistent,
mean field nonlinear Schr\"{o}dinger equation
 known as the  Gross-Pitaevskii equation \cite{Gross,Pit,Griffin}.
If a harmonic trap potential is included, the equation becomes
\be 
\label{gpe1}
i\hbar\pl{\psi(\bx,t)}{t}=-\fl{\hbar^2}{2m}\btd^2 \psi(\bx,t)+
\fl{m}{2}\left(\og_x^2 x^2+\og_y^2 y^2+\og_z^2 z^2\right)\psi(\bx,t)
+N U_0 |\psi(\bx,t)|^2\psi(\bx,t),
\ee 
where $\bx=(x,y,z)^T$ is the spatial coordinate vector, 
$m$ is the atomic mass, $\hbar=1.05\tm 10^{-34}\; [J\;s]$ 
is the Planck constant, 
$N$ is the number of atoms in the condensation,
and $\og_x$, $\og_y$ and $\og_z$ are the angular trap 
frequencies in $x$, $y$ and $z$-direction, respectively.
For the following we assume (w.r.o.g.) $\og_x\le \og_y\le \og_z$.
When $\og_x=\og_y=\og_z$, the trap potential is isotropic.
$U_0$ describes the interaction between atoms in the
condensation and has the form
\be
\label{U0}
U_0=\fl{4\pi \hbar^2 a}{m},
\ee
where $a$ is the $s$-wave scattering length (positive for repulsive
interaction and negative for attractive interaction). 
It is necessary to ensure that the wave
function is properly normalized. Specifically, we require
\be
\label{norm}
\int_{{\Bbb R}^3} \; |\psi(\bx,t)|^2\;d\bx=1.
\ee
A typical set of parameters used in current experiments with $^{87}$Rb is
\beas
&&m=1.44\tm 10^{-25}\ [kg], \
\og_x=\og_y=\og_z=20 \pi\ [1/s], \\ 
&&a=5.1\ [nm]=5.1\tm 10^{-9}\ [m], \ N:\ 10^2\sim 10^7.
\eeas

\subsection{Dimensionless GPE}

Following the physical literatures \cite{Edwards,Holland,Feder,
DalfoS,Tomio}, 
in order to scale the equation
(\ref{gpe1}) under the normalization 
(\ref{norm}),  we introduce 
\be
\label{dml}
\tilde{t}=\og_x t, \qquad \tilde{\bx}=\fl{\bx}{a_0}, \qquad
\tilde{\psi}(\tilde \bx,\tilde t)=a_0^{3/2} \psi(\bx,t),
\qquad \hbox{with}\qquad  a_0=\sqrt{\fl{\hbar}{\og_x m}},
\ee 
where $a_0$ is the length of the harmonic oscillator ground state.
In fact, here we choose $1/\og_x$ and $a_0$ as the dimensionless 
time and length units, respectively.
 Plugging (\ref{dml}) into (\ref{gpe1}), multiplying  by
$\fl{1}{m \og_x^2 a_0^{1/2}}$, 
 and then removing all \~{}, we
get the following dimensionless Gross-Pitaevskii equation
under the normalization (\ref{norm}) in three dimension
\be 
\label{gpe2}
i\;\pl{\psi(\bx,t)}{t}=-\fl{1}{2}\btd^2 \psi(\bx,t)+
V(\bx)\psi(\bx,t)
+ \kp\; |\psi(\bx,t)|^2\psi(\bx,t),
\ee 
where
\[
\label{poten}
V(\bx)=\fl{1}{2}\left(x^2+\gm_y^2 y^2+\gm_z^2 z^2\right),\quad 
\gm_y=\fl{\og_y}{\og_x}, \quad \gm_z=\fl{\og_z}{\og_x},\quad
\kp=\fl{U_0 N}{a_0^3\hbar \og_x}=\fl{4\pi aN}{a_0}.
\]
If we plug the typical set of parameters into above parameters, we find
the values
\[a_0\approx 0.3407\tm 10^{-5} [m], \qquad
\kp\approx 0.01881 N:\ 1.881\sim 188100.\]

   There are two extreme regimes: one is when  
$\kp=o(1)$, then equation (\ref{gpe2}) describes a 
weakly interacting  condensation. The other one is when
$\kp\gg1$, then (\ref{gpe2})
corresponds to a strongly interacting condensation or 
to the semiclassical regime.

\subsection{Reduction to lower dimension}

  In the following two cases, the 3d Gross-Pitaevskii equation (\ref{gpe2})
can approximately be reduced to 2d or even 1d \cite{Jackson, BeyondGPE}. 
In the case (disk-shaped condensation) 
\be
\label{r2d}
\og_x\approx \og_y, \quad \og_z\gg \og_x, \qquad \Longleftrightarrow
\quad \gm_y\approx1, \quad \gm_z\gg 1,
\ee
the 3d GPE (\ref{gpe2}) can be reduced to 2d  GPE with $\bx=(x,y)^T$ by 
assuming that the time evolution does not cause excitations along the 
$z$-axis since they have a large energy of approximately 
$\hbar \og_z$ compared to  
excitations along the $x$ and $y$-axis with energies of about 
$\hbar \og_x$. 
Thus we may assume that the condensation wave function along 
the $z$-axis is always 
well described by the ground state wave function and set \cite{Jackson,
BeyondGPE}
\be
\label{d2d}
\psi=\psi_2(x,y,t)\psi_3(z) \qquad \hbox{with}\quad
\psi_3(z)=\left(\int_{{\Bbb R}^2}\; |\phi_g(x,y,z)|^2 \;dxdy\right)^{1/2},
\ee
where $\phi_g(x,y,z)$ (see detail in (\ref{mp}))
is the ground state solution of the 3d GPE (\ref{gpe2}).
Plugging (\ref{d2d}) into (\ref{gpe2}), then multiplying by
$\psi^*_3(z)$ (where $f^*$ denotes the conjugate of a function $f$), 
integrating with respect to  $z$ over $(-\ift,\ift)$, we get
\be 
\label{gpe2dd}
i\;\pl{\psi_2(\bx,t)}{t}=-\fl{1}{2}\btd^2 \psi_2+
\fl{1}{2}\left(x^2+\gm_y^2 y^2+C\right) \psi_2
+ \left(\kp \int_{-\ift}^\ift \psi_3^4(z)\,dz\right) 
|\psi_2|^2\psi_2,
\ee 
where
\[C=\gm_z^2\int_{-\ift}^\ift \; z^2|\psi_3(z)|^2\;dz+ 
\int_{-\ift}^\ift\; \left|\fl{d\psi_3}{dz}\right|^2\;dz.\] 
Since this GPE is time-transverse invariant, we can replace 
$\psi_2\to \psi\;e^{-i\fl{C t}{2}}$ which  drops the constant $C$ in 
the trap potential  and obtain the  2d GPE, i.e.
\be 
\label{gpe2d}
i\;\pl{\psi(\bx,t)}{t}=-\fl{1}{2}\btd^2 \psi+
\fl{1}{2}\left(x^2+\gm_y^2 y^2\right) \psi
+ \left(\kp \int_{-\ift}^\ift \psi_3^4(z)\,dz\right) 
|\psi|^2\psi.
\ee 
The observables are not affected by this. 
\bigskip 

 In the case (cigar-shaped condensation) \cite{Jackson, BeyondGPE}
\be
\label{r2dd}
\og_y\gg \og_x, \quad \og_z\gg \og_x, \qquad \Longleftrightarrow \quad
\gm_y\gg1, \quad \gm_z\gg 1,
\ee
the 3d GPE (\ref{gpe2}) can be reduced to 1d  GPE with $\bx=x$. 
Similarly to the 2d case, we derive the 1d GPE \cite{Jackson, BeyondGPE}
\be 
\label{gpe1d}
i\; \pl{\psi(x,t)}{t}=-\fl{1}{2}\psi_{xx}(x,t)+
\fl{x^2}{2} \psi(x,t)
+ \left(\kp \int_{{\Bbb R}^2}\psi_{23}^4(y,z)\;dydz\right) \;
|\psi(x,t)|^2\psi(x,t),
\ee 
where 
\be
\label{psi23}
\psi_{23}(y,z)=\left(\int_{-\ift}^\ift\; \left|\phi_g(x,y,z)\right|^2
\;dx\right)^{1/2},
\ee
with $\phi_g(x,y,z)$ (see  (\ref{mp})) the ground state solution  
of (\ref{gpe2}).

   In fact, the 3d GPE (\ref{gpe2}), 2d GPE (\ref{gpe2d}) and
1d GPE (\ref{gpe1d}) can be written in a unified way 
\cite{BeyondGPE}
\be 
\label{gpeg}
i\;\pl{\psi(\bx,t)}{t}=-\fl{1}{2}\btd^2 \psi(\bx,t)+
V_d(\bx)\psi(\bx,t)
+ \kp_d\; |\psi(\bx,t)|^2\psi(\bx,t), \quad \bx\in {\Bbb R}^d,
\ee 
where
\be
\label{uf}
\kp_d=\dpm\left\{\ba{l}
\kp \ \dpm\int_{{\Bbb R}^2} \psi_{23}^4(y,z)\;dy dz, \\
\ \\
\kp \ \dpm\int_{-\ift}^\ift \psi_3^4(z)\;dz, \\
\ \\
\kp,\\
\ea\right.
\quad 
V_d(\bx)=\left\{\ba{ll}
 \dpm\fl{1}{2}x^2,  &\quad d=1, \\
\ \\
 \dpm\fl{1}{2}\left(x^2+\gm_y^2 y^2\right), &\quad d=2, \\
\ \\
 \dpm\fl{1}{2}\left(x^2+\gm_y^2 y^2+\gm_z^2 z^2\right), &\quad d=3.\\
\ea\right.
\ee
The normalization condition to (\ref{gpeg}) is
\be
\label{normg}
\int_{{\Bbb R}^d} \; |\psi(\bx,t)|^2\;d\bx=1.
\ee

\section{Ground state solution}\label{sna}  
\setcounter{equation}{0}  

In this section, we will propose a new numerical 
method by directly minimizing the energy functional
via finite element discretization
to compute the ground state solution of a BEC (\ref{gpeg}).
Furthermore we will also
provide approximate ground state solutions in two extreme
regimes: very weak interactions and strong repulsive interactions.

\subsection{Minimization problem}

To find a stationary solution of (\ref{gpeg}), we write 
\be
\label{stat}
\psi(\bx,t)=e^{-i\mu t}\; \phi(\bx),
\ee
where $\mu$ is the chemical potential
of the condensation and $\phi$ a real-valued function independent of time.
Inserting into (\ref{gpeg})
gives the following equation for $\phi(\bx)$
\be
\label{gss}
\mu \; \phi(\bx)=-\fl{1}{2}\btd^2 \phi(\bx)+
V_d(\bx)\phi(\bx)
+ \kp_d\; |\phi(\bx)|^2\phi(\bx), \qquad \bx\in {\Bbb R}^d,
\ee
under the normalization condition 
\be
\label{normgg}
\int_{{\Bbb R}^d} \; |\phi(\bx)|^2\;d\bx=1.
\ee
This is a nonlinear eigenvalue problem under a constraint and any eigenvalue
$\mu$ can be computed from its corresponding eigenfunction $\phi$ by
\be
\label{muphi}
\mu=\mub(\phi)=\int_{{\Bbb R}^d} \left[\fl{1}{2}
\left|\btd \phi(\bx)\right|^2+
V_d(\bx)\phi^2(\bx)
+\kp_d\;  \phi^4(\bx)\right]\;d\bx.
\ee
The Bose-Einstein condensation
ground-state wave function $\phi_g(\bx)$ is
found by solving this eigenvalue problem 
 under the normalization condition (\ref{normgg})
with the minimized chemical potential $\mu_g$. 
Usually, the ground state problem is formulated 
variationally.  Define the energy functional 
\begin{equation} 
\Eb(\phi):= \int_{{\Bbb R}^3} \left[\frac{1}{2}
\left|\nabla \phi\right|^2  
+ V_d(x) \left|\phi\right|^2  + \frac{\kappa}{2} 
\left|\phi\right|^4 \right]\;d\bx. 
\end{equation} 
It is easy to see that critical points of the energy functional
$\Eb(\phi)$ under the constraint 
(\ref{normgg}) are eigenfunctions of the 
nonlinear eigenvalue problem (\ref{gss}) under the constraint 
(\ref{normgg}) and versus versa. In fact, (\ref{gss}) can be viewed
as the Euler-Lagrange equation of the energy functional 
$\Eb(\phi)$  under the constraint 
(\ref{normgg}).
To compute the ground state $\phi_g$, we solve
the minimization problem

\bigskip

\noindent (V) Find $(\mu_g,\phi_g\in V)$ such that
\be
\label{mp}
\Eb(\phi_g)=\min_{\phi\in V}\; \Eb(\phi), \qquad \mu_g=
\mub(\phi_g)=\Eb(\phi_g)+
\int_{{\Bbb R}^d} \fl{\kp_d}{2}\; \phi_g^4(\bx)\;d\bx,
\ee
where the set $V$ is defined as
\[
V=\left\{\phi \ |\ \Eb(\phi)<\ift, \
\int_{{\Bbb R}^d} \; |\phi(\bx)|^2 \;d\bx=1\right\}.\]

  In non-rotating BEC, the minimization problem (\ref{mp}) 
 has a unique real valued nonnegative ground
state solution $\phi_g(\bx)>0$ for $\bx\in{\Bbb R}^d$
\cite{Lieb}. In physical literatures, the minimizer 
of (\ref{mp}) was obtained by either the 
Runge-Kutta type method \cite{Edwards,Adh,Adh1} 
or the imaginary time method \cite{Du, Tosi, Tosi2}. 
Here we present 
a method by directly minimizing the energy functional $\Eb(\phi)$
through the finite element discretization \cite{Ci}.

\subsection{Approximation in a bounded domain}

The eigenvalue problem (\ref{gss}) and the minimization problem
(\ref{mp}) are defined in ${\Bbb R}^d$. In practical
computation, usually they are approximated by problems defined
on a bounded computational domain. Since the full wave function must vanish
exponentially fast as $|\bx|\to \ift$ and due to symmetry,  
choosing $R_1, \cdots, R_d>0$ sufficiently large and denoting 
\[\Og^R=[0,R_1]\tm \cdots\tm [0,R_d],\] 
then the minimization problem
(\ref{mp}) can
be approximated  by

\bigskip

\noindent (V$^R$) Find $(\mu_g^R,\phi_g^R\in V_g)$ such that
\be
\label{mpb}
\Eb^R(\phi_g^R)=\min_{\phi\in V_g}\; \Eb^R(\phi), \qquad \mu_g^R=
\Eb^R(\phi_g^R)+
\int_{\Og^R} \fl{\kp_d}{2}\; \left[\phi_g^R(\bx)\right]^4\;d\bx,
\ee
where the set $V_g$ and the functional $\Eb^R(\phi)$ are defined as
\beas
&&V_g=\Biggl\{\phi\ |\ \Eb^R(\phi)<\ift, \ 
2^d \int_{\Og^R} \; |\phi(\bx)|^2\;d\bx=1,\ 
\phi(R_1,x_2,\cdots,x_d)=\cdots=
  \nn\\
&&\qquad \qquad \qquad \phi(x_1,\cdots,x_{d-1},R_d)=0\Biggr\}, 
  \nn\\
&&\Eb^R(\phi)=2^d\int_{\Og^R} \; \left[\fl{1}{2}
\left|\btd \phi(\bx)\right|^2+
V_d(\bx)\left|\phi(\bx)\right|^2
+\fl{\kp_d}{2}\; \left|\phi(\bx)\right|^4\right]\;d\bx.
\eeas

\subsection{Discretization}

The functional $\Eb^R(\phi)$ in (\ref{mpb}) (or $\Eb(\phi)$ in (\ref{mp}))
can be discretized by the finite element method \cite{Ci},
 finite difference method \cite{Morton} or
spectral method \cite{Gott}. Here we use the finite element method 
because it can easily keep the good properties of $\Eb^R(\phi)$, e.g.
positive, coercive and weakly lower semicontinuous when $\kp_d\ge0$,
on the unit sphere in finite dimensions. 
Let 
\[\tilde{V}_g=\left\{\phi\ |\ \Eb^R(\phi)<\ift, \ 
\phi(R_1,x_2,\cdots,x_d)=\cdots=\phi(x_1,\cdots,x_{d-1},R_d)=0\right\}\] 
and $\tilde{V}_g^h$ be a 
finite dimensional subspace of $\tilde{V}_g$ \cite{Ci}, i.e. 
$\tilde{V}_g^h\subset \tilde{V}_g$. Then the finite dimensional set
\[V_g^h = \left\{ \phi^h \in \tilde{V}_g^h\ |\ 
2^d \int_{\Og^R} \; |\phi^h(\bx)|^2\;d\bx=1,\right\}\]
is a subset of $V_g$, i.e.  $V_g^h\subset V_g$. Thus 
the minimization problem (\ref{mpb}) can 
be approximated  by

\bigskip

\noindent (V$^{R,h}$) Find $(\mu_g^{R,h},\phi_g^{R,h}\in V_g^h)$ such that
\be
\label{mpbht}
\Eb^R(\phi_g^{R,h})=\min_{\phi^h\in V_g^h}\; \Eb^R(\phi^h), 
\quad \mu_g^{R,h}=
\Eb^R(\phi_g^{R,h})+
\int_{\Og^R} \fl{\kp_d}{2}\; \left[\phi_g^{R,h}(\bx)\right]^4\;d\bx.
\ee

Introducing a functional with a Lagrange multiplier corresponding to the
normalization condition (\ref{normgg}), i.e.  
\beas
\tilde{E}_\kp^R(\phi^h,\ld)&=&\Eb^R(\phi^{h})-\ld \left( 
2^d \int_{\Og^R} \; |\phi^h(\bx)|^2\;d\bx-1\right)\\
&\equiv&F(\Phi^h,\ld), \qquad \phi^h\in \tilde{V}_g^h, \quad
 \ld \in {\Bbb R},
\eeas
where $\phi^h(\bx)=\Psi^h(\bx)\cdot \Phi^h \equiv 
\dpm\sum_{j=1}^M\; \Psi_1^h(\bx)
\Phi^h_j$ with $\Psi^h(\bx)=\left(\Psi_1^h(\bx), 
\cdots, \Psi_M^h(\bx)\right)^T$ a basis of the finite element subspace
$\tilde{V}_g^h$ and $\Phi^h=\left(\Phi_1^h, \cdots, \Phi_M^h\right)^T$
the unknowns of $\phi^h(\bx)$ \cite{Ci}, 
then the minimizer $\phi_g^{R,h}=\Psi^h(\bx)\cdot \Phi_g^{R,h}$ 
of the minimization problem
(\ref{mpbht}) is a critical point of the functional
$\tilde {E}_\kp^R(\phi^h,\ld)$. This implies
\be
\label{cp}
\left. \btd_{\Phi^h}\ F(\Phi^h,\ld)\right|_{ \Phi^h = \Phi_g^{R,h}}
={\bf 0},  \qquad \left.\pl{F(\Phi^h,\ld)}{\ld}
\right|_{ \Phi^h = \Phi_g^{R,h}}=0.
\ee
The nonlinear system (\ref{cp}) is solved by  the Newton's method \cite{Nash}
or quasi-Newton's method \cite{Nash} with a proper choice of the initial data 
$(\phi_g^{R,h})^{(0)}=\Psi^h(\bx)\cdot 
(\Phi_g^{R,h})^{(0)}$ and $\ld^{(0)}$ is the least square solution
of 
\[\left. \btd_{\Phi^h}\ F(\Phi^h,\ld)\right|_{\left(\Phi^h = 
(\Phi_g^{R,h})^{(0)},\ \ld=\ld^{(0)}\right) }
={\bf 0}. \]
The initial guess $(\phi_g^{R,h})^{(0)}$ is chosen as 
the interpoltant on $\tilde{V}_g^h$  of the
approximate ground state solution 
for very weak interactions  
(\ref{gssw1d}) or strong repulsive interactions (\ref{gssf1d}) 
when $\kp_d$ is not too big or
not too small, respectively. These approximate ground state solutions
are given in the next subsection.
Another way to choose the initial guess
is to use a continuation technique, i.e. 
use the numerical solution of the ground state function for a
 small $\kp_d$ as initial guess for computing the solution of a
larger $\kp_d$.

\subsection{Approximate ground state solution}

Here we present the approximate ground state solution of (\ref{gss})
in two extreme regimes: very weak interactions and strong 
repulsive interactions. 
These approximate ground state solutions are  used  as initial guess
$(\phi_g^{R,h})^{(0)}$ 
in our practical computation of the minimization problem 
(\ref{mpbht}) ( or (\ref{cp})).

   For a very weakly  interacting condensation, 
i.e. $\kp_d=o(1)$, we drop
the nonlinear term (i.e. the last term on the right hand side of 
(\ref{gss})) and get \cite{BeyondGPE}
\be 
\label{gpegw}
\mu\; \phi(\bx)=-\fl{1}{2}\btd^2 \phi(\bx)+
V_d(\bx)\;\phi(\bx), \qquad \bx\in {\Bbb R}^d.
\ee 
The ground state solution of (\ref{gpegw}) is
\be
\label{gsswww}
\mu_d^w=\fl{1}{2}\left\{\ba{l} 
1,\\ 
1+\gm_y,\\ 
1+\gm_y+\gm_z,\\ 
\ea\right. \  
\phi_d^w(\bx)=\fl{1}{(\pi)^{d/4}}\left\{\ba{ll} 
e^{-x^2/2}, & d=1,\\ 
\gm_y^{1/4}e^{-(x^2+\gm_y y^2)/2}, & d=2,\\ 
(\gm_y\gm_z)^{1/4}e^{-(x^2+\gm_y y^2+\gm_z z^2)/2}, & d=3.\\ 
\ea\right. 
\ee
This can be viewed as an approximate ground state solution of (\ref{gpeg})
in the case of a very weakly interacting 
 condensation.

   For strong repulsive interactions, i.e. $\kp_d\gg 1$, 
we drop the diffusion term (i.e. the first term on the right hand side
of (\ref{gss})) corresponding to the Thomas-Fermi approximation 
\cite{BeyondGPE}:
\be 
\label{gpeTF}
\mu\; \phi(\bx)=V_d(\bx)\phi(\bx)+\kp_d\; |\phi(\bx)|^2\phi(\bx), 
\qquad \bx\in {\Bbb R}^d,
\ee 
The ground state solution of (\ref{gpeTF}) is the compactly 
supported function $\phi_d^s(\bx)$:
\bea 
\label{gsskk} 
&&\mu_d^s=\left\{\ba{ll} 
\fl{1}{2}\left(\fl{3\kp_1 }{2}\right)^{2/3}, &\qquad \qquad d=1,\\ 
 \left(\fl{\kp_2 \gm_y}{\pi}\right)^{1/2},  &\qquad \qquad d=2,\\ 
\fl{1}{2}\left(\fl{15\kp_3\gm_y\gm_z}{4\pi}\right)^{2/5},  &\qquad \qquad
  d=3;\\ 
\ea\right. \\ 
\label{gssp} 
&&\phi_d^s(\bx)=\left\{\ba{ll} 
\sqrt{(\mu_d^s -V_d(\bx))/\kp_d}, &\qquad V_d(\bx)< \mu_d^s,\\ 
0, &\qquad  \hbox{otherwise}. 
\ea\right. 
\eea 

\begin{remark}
As indicated in Figure 6, an interface layer correction 
has to be constructed in order to improve the approximation quality
in the Thomas-Fermi regime (i.e. $\kp_d\gg1$).  
For a convergence proof of 
$\phi_3^s\to\phi_g$ as 
$\kp_3\to+\ift$ (without convergence rate) 
we refer to \cite{Lieb}. In Section \ref{sne},
these convergence rates are reported based on our 
numerical solutions.  
\end{remark}

\section{Ground state solution in 1d, 2d with radial symmetry and
3d with spherical symmetry}\label{s1d}  
\setcounter{equation}{0}  

In this section, we present detailed numerical formula and its 
finite element approximation for the ground state solution
of GPE in 1d, 2d with radial symmetry (i.e. $\gm_y=1$)
and 3d with spherical symmetry (i.e. $\gm_y=\gm_z=1$).
In these cases, the problem is reduced to 1d.
Due to symmetry, the GPE (\ref{gpeg})
essentially collapses to a 1d problem with $r=|\bx|\in[0,\ift)$ 
for $d=1,2,3$ 
\bea 
\label{gpeg1d}
&&i\;\pl{\psi(r,t)}{t}=-\fl{1}{2}\fl{1}{r^{d-1}}
\fl{\p}{\p r}\left(r^{d-1}\fl{\p}{\p r} \psi(r,t)\right)+
\fl{r^2}{2}\psi(r,t)
+ \kp_d\; |\psi(r,t)|^2\psi(r,t), \qquad  \\
\label{gpeg1dd}
&&\pl{\psi(0,t)}{r}=0, \qquad \psi(r,t)\to 0, \quad \hbox{when}\ 
r\to \ift. 
\eea 
The normalization condition collapses to
\be
\label{normg1d}
C_d \int_0^\ift \; |\psi(r,t)|^2\; r^{d-1}\;dr=1,
\ee
where
\[C_d=\left\{\ba{ll}
2, &\qquad d=1,    \\
2\pi, &\qquad d=2, \\
4\pi, &\qquad d=3.\\
\ea\right.
\]

\subsection{Minimization problem}

The eigenvalue problem (\ref{gss}) collapses to 
\bea
\label{gss1d}
&&\mu\;\phi(r)=-\fl{1}{2}\fl{1}{r^{d-1}}\fl{d}{dr}\left(r^{d-1}\fl{d}{dr}
 \phi(r)\right)+
\fl{r^2}{2}\phi(r)
+ \kp_d\; |\phi(r)|^2\phi(r), \ 0< r<\ift, \qquad \\
\label{gss1dd}
&&\phi^\prime(0)=0, \qquad \phi(r)\to 0, \quad \hbox{when}\ r\to \ift,
\eea 
under the normalization condition 
\be
\label{normgg1d}
C_d \int_0^\ift \; |\phi(r)|^2\; r^{d-1}\;dr=1.
\ee
The minimization problem (\ref{mp}) collapses to 

\bigskip

\noindent (V) Find $(\mu_g,\phi_g\in V)$ such that
\be
\label{mp1d}
\Eb(\phi_g)=\min_{\phi\in V}\; \Eb(\phi), \qquad \mu_g=\Eb(\phi_g)+
C_d \int_0^\ift \fl{\kp_d}{2}\; \phi_g^4(r)\;r^{d-1}\;dr,
\ee
where the set $V$ and the energy functional $\Eb(\phi)$ are defined as
\beas
&&V=\left\{\phi \ |\ \Eb(\phi)<\ift, \
C_d \int_0^\ift \; |\phi(r)|^2\; r^{d-1}\;dr=1\right\}, \nn\\
&&\Eb(\phi)=C_d\int_0^\ift \; \fl{r^{d-1}}{2}\left[
[\phi^\prime(r)]^2+
r^2\phi^2(r)
+ \kp_d\; \phi^4(r)\right]\;dr.
\eeas

\subsection{Approximation in a bounded domain}

The eigenvalue problem (\ref{gss1d}), (\ref{gss1dd}) 
and the minimization problem
(\ref{mp1d}) are defined in a semi-infinite interval $(0,\ift)$. 
In practical
computation, usually they are approximated by problems defined
on a finite interval. Since the full wave function must vanish
exponentially fast as $r\to \ift$,  choosing $R>0$ sufficiently large, then
the  eigenvalue problem (\ref{gss1d}), (\ref{gss1dd}) can
be approximated  by
\bea
\label{gss1db}
&&\mu\;\phi(r)=-\fl{1}{2}\fl{1}{r^{d-1}}\fl{d}{dr}\left(r^{d-1}\fl{d}{dr}
 \phi(r)\right)+
\fl{r^2}{2}\phi(r)
+ \kp_d\; |\phi(r)|^2\phi(r), \ 0< r<R, \qquad \\
\label{gss1dbb}
&&\phi^\prime(0)=0, \qquad \phi(R)=0,
\eea 
under the normalization condition 
\be
\label{normgg1dd}
C_d \int_0^R \; |\phi(r)|^2\; r^{d-1}\;dr=1.
\ee
Similarly the minimization problem (\ref{mp1d}) can be approximated by

\bigskip

\noindent (V$^R$) Find $(\mu_g^R,\phi_g^R\in V_g)$ such that
\be
\label{mpb1d}
\Eb^R(\phi_g^R)=\min_{\phi\in V_g}\; \Eb^R(\phi), \qquad 
\mu_g^R=\Eb^R(\phi_g^R)+
C_d \int_0^R \fl{\kp_d}{2}\; \left[\phi_g^R(r)\right]^4\; r^{d-1} \;dr,
\ee
where the set $V_g$ and the functional $\Eb^R(\phi)$ are defined as
\beas
&&V_g=\left\{\phi\ |\ \Eb^R(\phi)<\ift, \ 
C_d \int_0^R \; |\phi(r)|^2\; r^{d-1}\;dr=1,\ \phi(R)=0\right\}, 
  \nn\\
&&\Eb^R(\phi)=C_d\int_0^R \;\fl{r^{d-1}}{2} \left[
[\phi^\prime(r)]^2+
r^2\phi^2(r)
+ \kp_d\; \phi^4(r)\right]\;dr.
\eeas

\subsection{Finite element approximation}

Assume that                
\[0=r_0<r_1<r_2<\cdots<r_M=R\]
is a partition of the interval $[0,R]$ with mesh size $h$ \cite{Ci}. Let
\beas
\tilde{V}_g^h&=&\Bigl\{\phi^h(r)\in C([0,R])\ |\  
\phi^h(r)|_{[r_j,r_{j+1}]}\in 
P_1([r_j,r_{j+1}]),\quad 0\le j\le M-1,\\
&&\qquad \phi^h(R)=0\Bigr\}, \\
V_g^h&=&\left\{\phi^h\in \tilde{V}_g^h 
\ |\  C_d \int_0^R \; |\phi^h(r)|^2\; r^{d-1}\;dr=1,\right\}; 
\eeas
where $P_1$ denotes piecewise linear polynomials. 
Then the finite element approximation of the problem (\ref{mp1d}) is

\bigskip

\noindent (V$^{R,h}$) Find $(\mu_g^{R,h},\phi_g^{R,h}\in V_g^h)$ such that
\be
\label{mpbh}
\Eb^R(\phi_g^{R,h})=\min_{\phi\in V_g^h}\; \Eb^R(\phi), 
\quad \mu_g^{R,h}=\Eb^R(\phi_g^{R,h})+
C_d \int_0^R \fl{\kp_d}{2}\; \left[\phi_g^{R,h}(r)\right]^4\; r^{d-1} \;dr.
\ee

\subsection{Approximate ground state solution}

In these cases, the approximate ground state solution collapses to
the following.
   For a very weakly  interacting condensation, 
i.e. $\kp_d=o(1)$,
the ground state solution is
\be
\label{gssw1d}
\mu_d^w=\fl{d}{2}, \qquad 
\phi_d^w(r)=\fl{1}{\pi^{d/4}}
\; e^{-r^2/2}, \qquad d=1,2,3.
\ee
For strong repulsive interactions, i.e. $\kp_d\gg 1$, 
the ground state solution is
\bea
\label{gss1dr}
&&\mu_d^s=\fl{1}{2}\left[\fl{((d+1)^2-1)\kp_d}{C_d}\right]^{2/(d+2)}, 
\qquad 
d=1,2,3,  \\
\label{gssf1d}
&&\phi_d^s(r)=\left\{\ba{ll}
\sqrt{\left(\mu_g^s  -r^2/2\right)/\kp_d}, 
&\qquad  r^2< 2\mu_d^s,\\
0, &\qquad  \hbox{otherwise},
\ea\right.
\qquad  d=1,2,3.
\eea

\section{Ground state solution in 3d with cylindrical 
symmetry}\label{s2d}  
\setcounter{equation}{0}  

In this section, we present detailed numerical formula and its 
finite element approximation for the ground state solution
of GPE in  3d with cylindrical symmetry (i.e. $\gm_y=1$).
In this case, the problem is reduced to 2d.
Due to symmetry, the GPE (\ref{gpeg})
essentially collapses to a 2d problem with $r=\sqrt{x^2+y^2}\in[0,\ift)$ 
\bea 
\label{gpeg2d}
&&i\;\pl{\psi(r,z,t)}{t}=-\fl{1}{2}\left[\fl{1}{r}
\fl{\p}{\p r}\left(r\fl{\p \psi}{\p r}\right)+\fl{\p^2\psi}{\p z^2}\right]+
\fl{1}{2}(r^2+\gm_z^2 z^2)\psi
+ \kp\; |\psi|^2\psi, \\
&&\qquad \qquad \qquad \qquad \qquad \qquad  0<r,z<\ift, \nn \\
\label{gpeg2dd}
&&\pl{\psi(0,z,t)}{r}=0,\quad 0\le z<\ift, \qquad 
 \pl{\psi(r,0,t)}{z}=0, \quad 0\le r<\ift, \\
\label{gpeg2dk}
&&\psi(r,z,t)\to 0, \quad \hbox{when}\ r+|z|\to \ift. 
\eea 
The normalization condition collapses to
\be
\label{normg2d}
4\pi \int_0^\ift\int_0^\ift \; |\psi(r,z,t)|^2\; r\;drdz=1.
\ee

\subsection{Minimization problem}

The eigenvalue problem (\ref{gss}) collapses to 
\bea
\label{gss2d}
&&\mu\;\phi(r,z)=-\fl{1}{2}\left[\fl{1}{r}
\fl{\p}{\p r}\left(r\fl{\p \phi}{\p r}\right)+\fl{\p^2\phi}{\p z^2}\right]+
\fl{1}{2}(r^2+\gm_z^2 z^2)\phi
+ \kp\; |\phi|^2\phi, \\
&&\qquad \qquad \qquad  \qquad \qquad \qquad \ 0< r,z<\ift, \nn \\
\label{gss2dd}
&&\pl{\phi(0,z)}{r}=0,\quad 0\le z<\ift, \qquad 
 \pl{\phi(r,0)}{z}=0, \quad 0\le r<\ift, \\
\label{gss2dk}
&&\phi(r,z)\to 0, \quad \hbox{when}\ r+|z|\to \ift, 
\eea 
under the normalization condition 
\be
\label{normgg2d}
4\pi \int_0^\ift\int_0^\ift \; |\phi(r,z)|^2\; r\;drdz=1.
\ee
The minimization problem (\ref{mp}) collapses to 

\bigskip

\noindent (V) Find $(\mu_g,\phi_g\in V)$ such that
\be
\label{mp2d}
\Eb(\phi_g)=\min_{\phi\in V}\; \Eb(\phi), \quad \mu_g=\Eb(\phi_g)+
4\pi \int_0^\ift\int_0^\ift \fl{\kp}{2}\; \phi_g^4(r,z)\;r\;drdz,
\ee
where the set $V$ and the functional $\Eb(\phi)$ are defined as
\beas
&&V=\left\{\phi \ |\ \Eb(\phi)<\ift, \
4\pi \int_0^\ift\int_0^\ift \; |\phi(r,z)|^2\; r\;drdz=1\right\}, \nn\\
&&\Eb(\phi)=4\pi\int_0^\ift\int_0^\ift \; 
\fl{r}{2}\left[\phi_r^2(r,z)+\phi_z^2(r,z)+
(r^2+\gm_z^2 z^2)\phi^2(r,z)
+ \kp\;\phi^4(r,z)\right]\;drdz.
\eeas

\subsection{Approximation in a bounded domain}

The eigenvalue problem (\ref{gss2d})-(\ref{gss2dk}) 
and the minimization problem
(\ref{mp2d}) are defined in the first quadrant of the $rz$-plane. 
In practical computation, usually they are approximated by problems defined
on a bounded domain. Since the full wave function must vanish
exponentially fast as $r+|z|\to \ift$,  choosing $R>0$ and $Z>0$ 
sufficiently large, then
the  eigenvalue problem (\ref{gss2d})-(\ref{gss2dk}) can
be approximated  by
\bea
\label{gss2db}
&&\mu\;\phi(r,z)=-\fl{1}{2}\left[\fl{1}{r}
\fl{\p}{\p r}\left(r\fl{\p \phi}{\p r}\right)+\fl{\p^2\phi}{\p z^2}\right]+
\fl{1}{2}(r^2+\gm_z^2 z^2)\phi
+ \kp\; |\phi|^2\phi, \nn\\
&&\qquad \qquad \qquad \qquad \qquad \qquad 
\qquad \quad 0< r<R, \ 0<z<Z, \qquad \\
\label{gss2ddd}
&&\pl{\phi(0,z)}{r}=0,\quad 0\le z\le Z, \qquad 
 \pl{\phi(r,0)}{z}=0, \quad 0\le r \le R, \\
\label{gss2dbb}
&&\phi(R,z)=0, \quad 0\le z\le Z, \qquad \phi(r,Z)=0, \quad 0\le r\le R,
\eea 
under the normalization condition 
\be
\label{normgg2dd}
4\pi \int_0^R\int_0^Z \; |\phi(r,z)|^2\; r\;dzdr=1.
\ee
Similarly the minimization problem (\ref{mp2d}) can be approximated by

\bigskip

\noindent (V$^R$) Find $(\mu_g^R,\phi_g^R\in V_g)$ such that
\be
\label{mp2db}
\Eb^R(\phi_g^R)=\min_{\phi\in V_g}\; \Eb^R(\phi), \quad \mu_g^R=
\Eb^R(\phi_g^R)+
4\pi \int_0^R\int_0^Z \fl{\kp}{2}\; \left[\phi_g^R(r,z)\right]^4
\;r\; drdz,
\ee
where the set $V_g$ and the functional $J^R(\phi)$ are defined as
\beas
&&V_g=\Biggl\{\phi\ |\ \Eb^R(\phi)<\ift, \ 
4\pi \int_0^R\int_0^Z \; |\phi(r,z)|^2\; r\;dzdr=1,\\
&&\qquad \qquad  \phi(R,z)=0, \ 0\le z\le Z, \quad \phi(r,Z)=
0, \ 0\le r\le R\Biggr\}, \\
&&\Eb^R(\phi)=4\pi\int_0^R\int_0^Z \; 
\fl{r}{2}\left[\phi_r^2(r,z)+\phi_z^2(r,z)+
(r^2+\gm_z^2 z^2)\phi^2(r,z)
+ \kp\;\phi^4(r,z)\right]\;dzdr.
\eeas

\subsection{Finite element approximation}

Assume that
\[0=r_0<r_1<r_2<\cdots<r_M=R, \quad 0=z_0<z_1<z_2<\cdots<Z_N=Z\]
is a partition of the rectangle $[0,R]\tm[0,Z]$ 
with mesh size $h$ \cite{Ci}. Let
\beas
\tilde{V}_g^h&=&\Biggl\{\phi^h(r,z)\in C([0,R]\tm[0,Z])\ |\  
\phi^h(r,z)|_{[r_j,r_{j+1}]\tm[z_l,z_{l+1}]}\in 
Q_1([r_j,r_{j+1}]\tm[z_l,z_{l+1}]),\\
&&\qquad 0\le j\le M-1,\ 0\le l\le N-1,\ 
 \phi^h(R,z)=0,\ 0\le z\le Z, \\
&& \qquad \quad  \phi^h(r,Z)=0, \  0\le r\le R\Biggr\},\\
V_g^h&=&\left\{\phi^h \in \tilde{V}_g^h\ |\
 4\pi \int_0^R\int_0^Z \; |\phi^h(r,z)|^2\; r\;dzdr=1\right\};
\eeas
where $Q_1$ denotes all bilinear polynomials.
Then the finite element approximation of the problem (\ref{mp2db}) is

\bigskip

\noindent (V$^{R,h}$) Find $(\mu_g^{R,h},\phi_g^{R,h}\in V_g^h)$ such that
\be
\label{mpbh2d}
\Eb^R(\phi_g^{R,h})=\min_{\phi\in V_g^h}\; \Eb^R(\phi), \ \mu_g^{R,h}=
\Eb^R(\phi_g^{R,h})+
4\pi \int_0^R\int_0^Z \fl{\kp}{2}\; \left[\phi_g^{R,h}(r,z)\right]^4
\;r\;drdz.
\ee

\subsection{Approximate ground state solution}

In this case, the approximate ground state solution collapses to
the following.
   For a very weakly  interacting condensation, 
i.e. $\kp_3=\kp=o(1)$,
the ground state solution  is
\be
\label{gssw2d}
\mu_3^w=1+\fl{\gm_z}{2}, \qquad 
\phi_3^w(r,z)=\fl{\gm_z^{1/4}}{(\pi)^{3/4}}
e^{-(r^2+\gm_z z^2)/2}.
\ee
For strong repulsive interactions, i.e. $\kp_3=\kp\gg 1$, 
the ground state solution is
\bea
\label{gss2dr}
&&\mu_3^s=\fl{1}{2}\left(\fl{15\kp\gm_z}{4\pi}\right)^{2/5},\\
\label{gssp2d} 
&&\phi_3^s(r,z)=\left\{\ba{ll}
\sqrt{(\mu_3^s -(r^2+\gm_z^2 z^2)/2)/\kp}, 
  &\qquad r^2+\gm_z^2 z^2< 2\mu_3^s,\\ 
0, &\qquad \hbox{otherwise}. \\
\ea\right. 
\eea
 
\section{Numerical results}\label{sne}  
\setcounter{equation}{0}  
  
  In this section we shall report on numerical error analysis
of our method, numerical ground state
solutions of (\ref{gpeg}) in 1d, 2d with radial 
symmetry and 3d with spherical
symmetry as well as cylindrical symmetry. Furthermore we also 
compare the numerical ground state solution of (\ref{gpeg}) in 3d and 
the corresponding Thomas-Fermi approximation (\ref{gssp2d}).
Convergence rates of the Thomas-Fermi approximations to their
exact counterparts are also reported.

\subsection{Numerical error analysis}

   In this subsection, we study numerically the convergence rate of 
the finite element discretization to the minimization problem 
(\ref{mpb}) in 1d.
We choose $d=1$ and $\kp_d=\kp_1=62.742$ in (\ref{mpb1d}).
We compute a numerical solution of (\ref{mpb1d}) in 1d on 
$\Og=[0,8]$ by using the discretization (\ref{mpbh}) with a very fine
mesh, e.g. $h=\fl{1}{128}$, as the ``exact'' ground state 
solution $\phi_g(x)$.   Table 1 shows the errors 
$\Eb(\phi_g^h)-\Eb(\phi_g)$, $\mu_g^h-\mu_g$,  
$\max\left|\phi_g^h(x) -\phi_g(x)\right|$, 
$\left\|\phi_g^h -\phi_g\right\|_{L^2(\Og)}$,
$\left\|(\phi_g^h)^2 -(\phi_g)^2\right\|_{L^1(\Og)}$ and
 $\left\|\phi_g^h -\phi_g\right\|_{H^1(\Og)}$ for different 
mesh size $h$. Here we use the standard Sobolev space norms \cite{Adm}.

\begin{table}[htbp]
\begin{center}
\begin{tabular}{cccccc}\hline
\   &$h=\fl{1}{2}$   &$h=\fl{1}{4}$  &$h=\fl{1}{8}$ 
&$h=\fl{1}{16}$ &$h=\fl{1}{32}$  \\ \hline \\
$\Eb(\phi_g^h)-\Eb(\phi_g)$  
  &6.528E-4 &1.570E-4 &3.878E-5 &9.560E-6 &2.274E-6\\ 
$\mu_g^h-\mu_g$  
  &3.462E-4 &8.073E-5 &1.986E-5 &4.887E-6 &1.160E-6\\ 
$\max\left|\phi_g^h(x)-\phi_g(x)\right|$  
  &3.222E-3 &8.450E-4 &2.091E-4 &5.300E-5 &1.359E-5\\ 
$\left\|\phi_g^h -\phi_g\right\|_{L^2(\Og)}$ 
  &2.177E-3 &5.290E-4 &1.323E-4 &3.394E-5  &9.295E-6\\ 
$\left\|(\phi_g^h)^2 -(\phi_g)^2\right\|_{L^1(\Og)}$ 
  &1.042E-3 &2.633E-4 &6.579E-5 &1.700E-5 &4.927E-6\\ 
$\left\|\phi_g^h -\phi_g\right\|_{H^1(\Og)}$ 
  &2.493E-2 &1.246E-2 &6.218E-3 &3.091E-3 &1.508E-3\\
\\ \hline

\end{tabular}
\end{center}

Table 1. Numerical error analysis of the finite 
element discretization (\ref{mpbh}). 
\end{table}

  From Table 1, we observe that the approximate 
energy $\Eb(\phi_g^h)$, chemical potential $\mu_g^h$,
ground state solution $\phi_g^h$, and atom density
 function $(\phi_g^h)^2$
 converge to $\Eb(\phi_g)$, $\mu_g$, $\phi_g$ in maximum norm
and $L^2$-norm, and $(\phi_g)^2$ in $L^1$-norm, respectively, 
at second order convergence rate when the mesh size $h$ goes to zero.
Furthermore $\phi_g^h$ converges to $\phi_g$
in $H^1$-norm at first order convergence rate.

\subsection{Results in 1d, 2d with radial symmetry and
3d with spherical symmetry}

An interesting property of the condensation 
wave function in these cases is its root mean square (rms) size 
$r_{\rm rms}$ defined by
\be
\label{msr}
r_{\rm rms}^2=\langle r^2\rangle=C_d\int_0^\ift r^2\; 
\phi^2(r)\;r^{d-1}\;dr, 
\qquad d=1,2,3.
\ee
In our computations, we choose $R=16$ in (\ref{gss1db}) with a uniform
partition of the interval $[0,R]$ of mesh size $h=\fl{1}{50}$ 
in (\ref{mpbh}).

   Figure 1 shows the ground-state condensation wave function $\phi_g(r)$ 
(with $\phi_g(-r)=\phi_g(r)$ for $r\in{\Bbb R}$ in 1d) 
versus $r$ and the chemical potential $\mu_g$ for different $\kp_d$,
  and Table 2 lists $\phi_g(0)$, $r_{\rm rms}$, 
$\mu_g$ versus $\kp_d$ for $d=1$. For comparison, we also list the
chemical potential $\mu_g$ obtained 
 by the Thomas-Femi approximation 
(TFA) in (\ref{gss1dr}).
Furthermore Figure 2 and Table 3 show similar results for 
$d=2$, and Figures 3 and Table 4  for $d=3$.
 
\begin{table}[htbp]
\begin{center}
\begin{tabular}{cccc}\hline
$\kp_1$  &$\phi_g(0)$ &$r_{\rm rms}$ &$\mu_g$ \\
\ &\ &\ &$\ba{cc}
\hbox{Numerical} &\hbox{TFA}\\
\ea$ \\ \hline
-12.5484    &1.7718     &0.1444   &$\ba{cc}
\quad -19.669 &\quad \hbox{NA\ }\ea$ \\
-6.2742    &1.2654     &0.2810   &$\ba{cc}
\quad -4.9553 &\quad \hbox{NA\ }\ea$ \\
-2.5097    &0.9132     &0.5133   &$\ba{cc}
\quad -0.8061 &\quad \hbox{NA\ }\ea$ \\
 0   &0.7511     &0.7071   &$\ba{cc}
\quad 0.5000 &\quad \hbox{NA\ }\ea$ \\
3.1371    &0.6459     &0.8960   &$\ba{cc}
\quad 1.5265 &\quad \hbox{NA\ }\ea$ \\
 12.5484   &0.5297     &1.2454   &$\ba{cc}
\quad 3.5965  &\quad 3.538\ea$ \\
 31.371   &0.4556     &1.6416   &$\ba{cc}
\quad 6.5526 &\quad 6.517\ea$ \\
 62.742  &0.4060     &2.0495   &$\ba{cc}
\quad 10.369  &\quad 10.345 \ea$ \\
 156.855   &0.3485     &2.7679   &$\ba{cc}
\quad 19.0704 &\quad 19.056\ea$ \\
 313.71   &0.3105     &3.4823   &$\ba{cc}
\quad 30.259 &\quad 30.249\ea$ \\
 627.42  &0.2766     &4.3847   &$\ba{cc}
\quad  48.024 &\quad 48.018\ea$ \\
 1254.8   &0.2464     &5.5228   &$\ba{cc}
\quad 76.226 &\quad 76.222\ea$ \\
   \hline
\end{tabular}
\end{center}

Table 2. Ground state chemical potential $\mu_g$, maximum value 
of the wave function
$\phi_g(0)$ and root mean square size $r_{\rm rms}$
versus the interaction coefficient $\kp_d$
in 1d ($d=1$). 
\end{table}

\begin{table}[htbp]
\begin{center}
\begin{tabular}{cccc}\hline
$\kp_2$  &$\phi_g(0)$ &$r_{\rm rms}$ &$\mu_g$ \\
\ &\ &\ &$\ba{cc}
\hbox{Numerical} &\hbox{TFA}\\
\ea$ \\ \hline
-5.8    &2.1770     &0.3208   &$\ba{cc}
\quad -5.552  &\quad \hbox{NA\ }\ea$ \\
-4.5    &0.8948     &0.7091   &$\ba{cc}
\quad -0.2923  &\quad \hbox{NA\ }\ea$ \\
-2.5097    &0.6754     &0.8775   &$\ba{cc}
\quad 0.4997 &\quad \hbox{NA\ }\ea$ \\
 0   &0.5642     &1.0000   &$\ba{cc}
\quad 1.0000 &\quad \hbox{NA\ }\ea$ \\
 3.1371   &0.4913     &1.1051   &$\ba{cc}
\quad 1.4200 &\quad \hbox{NA\ }\ea$ \\
12.5484    & 0.3919    &1.3068   &$\ba{cc}
\quad  2.2558 &\quad 1.9986 \ea$ \\
  62.742  &0.2676     &1.7881   &$\ba{cc}
\quad 4.6098 &\quad 4.4689\ea$ \\
 313.71   &0.1787     &2.6044   &$\ba{cc}
\quad 10.068  &\quad  9.9928\ea$ \\
627.42    &0.1502     &3.0845   &$\ba{cc}
\quad  14.1892 &\quad 14.132 \ea$ \\
 1254.8   &0.1262     &3.6598   &$\ba{cc}
\quad 20.0286  &\quad 19.9854\ea$ \\ 
 \hline
\end{tabular}
\end{center}
Table 3. Ground state chemical potential $\mu_g$, maximum value 
of the wave function
$\phi_g(0)$ and root mean square size $r_{\rm rms}$
versus the interaction coefficient $\kp_d$
in 2d with radial symmetry ($d=2$). 
\end{table}

\begin{table}[htbp]
\begin{center}
\begin{tabular}{cccc}\hline
$\kp_3$  &$\phi_g(0)$ &$r_{\rm rms}$ &$\mu_g$ \\
\ &\ &\ &$\ba{cc}
\hbox{Numerical} &\hbox{TFA}\\
\ea$ \\ \hline
 -7   &0.7613     &0.9512   &$\ba{cc}
\quad 0.6210  &\quad \hbox{NA\ }\ea$ \\
 -3.1371   &0.4881     &1.1521   &$\ba{cc}
\quad 1.2652  &\quad \hbox{NA\ }\ea$ \\
0    &0.4238     &1.2248   &$\ba{cc}
\quad 1.5000  &\quad \hbox{NA\ }\ea$ \\
3.1371    &0.3843     &1.2785   &$\ba{cc}
\quad 1.6774  &\quad \hbox{NA\ }\ea$ \\
12.5484    &0.3180     &1.3921   &$\ba{cc}
\quad 2.0650  &\quad 1.4762\ea$ \\
31.371    &0.2581     &1.5356   &$\ba{cc}
\quad 2.5861  &\quad 2.1298\ea$ \\
125.484    &0.1738     &1.8821   &$\ba{cc}
\quad 4.0141  &\quad 3.7082\ea$ \\
627.4    &0.1066     &2.5057   &$\ba{cc}
\quad 7.2484  &\quad 7.059\ea$ \\
3137.1    &0.0655     &3.4145   &$\ba{cc}
\quad 13.553  &\quad 13.438\ea$ \\
31371    &0.0328     &5.3852   &$\ba{cc}
\quad 33.810  &\quad 33.755\ea$ \\
 \hline
\end{tabular}
\end{center}
Table 4. Ground state chemical potential $\mu_g$, maximum value 
of the wave function
$\phi_g(0)$ and root mean square size $r_{\rm rms}$
versus the interaction coefficient $\kp_d$
in 3d with spherical symmetry ($d=3$). 
\end{table}

\bigskip

  From Figures 1-3 and Tables 2-4, we can see that the chemical 
potential $\mu_g$ and the root mean square size  will increase when 
the interaction coefficient $\kp_d$ ( i.e. the number of atoms in
the condensation) is increasing. On the other 
hand, the peak of the ground state solution $\phi_g(0)$ will decrease.
If we use the typical set of parameter values in Section 2, a $\kp_3=
31371$ corresponds to a condensation population of $N\approx 1\;667\;800$,
i.e. approximately one and a half millions atoms in the 
condensation. Furthermore the Thomas-Femi approximation gives accurate 
chemical potential $\mu_g$ only when the interaction $\kp_d$ is very big
and poor approximation for intermediate values of $\kp_d$ 
(cf. Tables 2-4). 

  Here we also report the convergence rates of the 
Thomas-Femi approximations $\phi_d^s$ and $\mu_d^s$
to the exact ground state solution $\phi_g$ and $\mu_g$,
respectively,  as $\fl{1}{\kp_d}\to 0$. Tables 5,6\&7 list
these results for $d=1,2\&3$, respectively.

\begin{table}[htbp]
\begin{center}
\begin{tabular}{ccccccc}\hline
\\ 
     &$\dpm\fl{1}{\kp_1}$   &$\dpm\fl{1}{100}$    &$\dpm\fl{1}{200}$ 
&$\dpm\fl{1}{400}$ &$\dpm\fl{1}{800}$   &$\dpm\fl{1}{1600}$\\
 \\
\hline
$\left|\mu_g-\mu_1^s\right|$  
&\begin{tabular}{c}
Error \\
Rate \\
\end{tabular} & 
\begin{tabular}{c}
1.875E-2 \\
0.5655 \\
\end{tabular} & 
\begin{tabular}{c}
1.267E-2 \\
 0.5711\\
\end{tabular} & 
\begin{tabular}{c}
8.528E-3 \\
0.5775 \\
\end{tabular} & 
\begin{tabular}{c}
5.715E-3  \\
 0.5827  \\
\end{tabular} & 
\begin{tabular}{c}
3.816E-3  \\
   \\
\end{tabular} \\ \\

$\left\|\phi_g^2-(\phi_1^s)^2\right\|_{L^1}$  
&\begin{tabular}{c}
Error \\
Rate \\
\end{tabular} & 
\begin{tabular}{c}
9.915E-3 \\
0.7738 \\
\end{tabular} & 
\begin{tabular}{c}
5.799E-3 \\
 0.7488\\
\end{tabular} & 
\begin{tabular}{c}
3.451E-3 \\
0.7151 \\
\end{tabular} & 
\begin{tabular}{c}
2.102E-3  \\
 0.6723  \\
\end{tabular} & 
\begin{tabular}{c}
1.319E-3  \\
   \\
\end{tabular} \\ \\

$\left\|\phi_g-\phi_1^s\right\|_{L^2}$  
&\begin{tabular}{c}
Error \\
Rate \\
\end{tabular} & 
\begin{tabular}{c}
5.709E-2 \\
0.4346 \\
\end{tabular} & 
\begin{tabular}{c}
4.224E-2 \\
 0.4371\\
\end{tabular} & 
\begin{tabular}{c}
3.120E-2 \\
0.4368 \\
\end{tabular} & 
\begin{tabular}{c}
2.305E-2  \\
 0.4375  \\
\end{tabular} & 
\begin{tabular}{c}
1.702E-2  \\
   \\
\end{tabular} \\ \\

$\left\|\phi_g-\phi_1^s\right\|_{L^4}$  
&\begin{tabular}{c}
Error \\
Rate \\
\end{tabular} & 
\begin{tabular}{c}
5.676E-2 \\
0.4079 \\
\end{tabular} & 
\begin{tabular}{c}
4.278E-2 \\
 0.4103\\
\end{tabular} & 
\begin{tabular}{c}
3.219E-2 \\
0.4080 \\
\end{tabular} & 
\begin{tabular}{c}
2.426E-2  \\
 0.4083  \\
\end{tabular} & 
\begin{tabular}{c}
1.828E-2  \\
   \\
\end{tabular} \\ \\

\hline 
\end{tabular}  
\end{center}

Table 5. Convergence rates of the Thomas-Femi approximations
as $\dpm\fl{1}{\kp_1}\to 0$ in 1d ($d=1$). 
\end{table}

\begin{table}[htbp]
\begin{center}
\begin{tabular}{ccccccc}\hline
\\ 
     &$\dpm\fl{1}{\kp_2}$  &$\dpm\fl{1}{200}$ &$\dpm\fl{1}{400}$    
&$\dpm\fl{1}{800}$  &$\dpm\fl{1}{1600}$ &$\dpm\fl{1}{3200}$   \\
 \\
\hline
$\left|\mu_g-\mu_2^s\right|$  
&\begin{tabular}{c}
Error \\
Rate \\
\end{tabular} & 
\begin{tabular}{c}
9.004E-2  \\
 0.3947  \\
\end{tabular}& 
\begin{tabular}{c}
6.849E-2 \\
0.4007 \\
\end{tabular} & 
\begin{tabular}{c}
5.188E-2 \\
 0.4062\\
\end{tabular} & 
\begin{tabular}{c}
3.915E-2 \\
0.4112 \\
\end{tabular} & 
\begin{tabular}{c}
2.944E-2  \\
   \\
\end{tabular}  \\ \\

$\left\|\phi_g^2-(\phi_2^s)^2\right\|_{L^1}$  
&\begin{tabular}{c}
Error \\
Rate \\
\end{tabular} & 
\begin{tabular}{c}
4.458E-2  \\
 0.5835  \\
\end{tabular}& 
\begin{tabular}{c}
2.975E-2 \\
0.5881 \\
\end{tabular} & 
\begin{tabular}{c}
1.979E-2 \\
 0.5941\\
\end{tabular} & 
\begin{tabular}{c}
1.311E-2 \\
0.5828 \\
\end{tabular} & 
\begin{tabular}{c}
8.753E-3  \\
   \\
\end{tabular}  \\ \\

$\left\|\phi_g-\phi_2^s\right\|_{L^2}$  
&\begin{tabular}{c}
Error \\
Rate \\
\end{tabular} & 
\begin{tabular}{c}
1.292E-1 \\
0.3158 \\
\end{tabular} & 
\begin{tabular}{c}
1.038E-1 \\
 0.3164\\
\end{tabular} & 
\begin{tabular}{c}
8.336E-2 \\
0.3182 \\
\end{tabular} & 
\begin{tabular}{c}
6.686E-2  \\
 0.3211  \\
\end{tabular} & 
\begin{tabular}{c}
5.352E-2  \\
   \\
\end{tabular} \\ \\

$\left\|\phi_g-\phi_2^s\right\|_{L^4}$  
&\begin{tabular}{c}
Error \\
Rate \\
\end{tabular}  & 
\begin{tabular}{c}
6.613E-2  \\
 0.3554  \\
\end{tabular} & 
\begin{tabular}{c}
5.169E-2 \\
 0.3570\\
\end{tabular} & 
\begin{tabular}{c}
4.036E-2 \\
 0.3580\\
\end{tabular} & 
\begin{tabular}{c}
3.149E-2 \\
0.3627 \\
\end{tabular} & 
\begin{tabular}{c}
2.449E-2  \\
   \\
\end{tabular} \\ \\

\hline 
\end{tabular}  
\end{center}

Table 6. Convergence rates of the Thomas-Femi approximations
as $\dpm\fl{1}{\kp_2}\to 0$ in 2d ($d=2$). 
\end{table}

\begin{table}[htbp]
\begin{center}
\begin{tabular}{ccccccc}\hline
\\ 
     &$\dpm\fl{1}{\kp_3}$  &$\dpm\fl{1}{400}$ &$\dpm\fl{1}{800}$    
&$\dpm\fl{1}{1600}$  &$\dpm\fl{1}{3200}$ &$\dpm\fl{1}{6400}$   \\
 \\
\hline
$\left|\mu_g-\mu_3^s\right|$  
&\begin{tabular}{c}
Error \\
Rate \\
\end{tabular} & 
\begin{tabular}{c}
2.169E-1  \\
 0.3023  \\
\end{tabular}& 
\begin{tabular}{c}
1.759E-1 \\
0.3058 \\
\end{tabular} & 
\begin{tabular}{c}
1.423E-1 \\
 0.3111\\
\end{tabular} & 
\begin{tabular}{c}
1.147E-1 \\
0.3142 \\
\end{tabular} & 
\begin{tabular}{c}
9.225E-2  \\
   \\
\end{tabular}  \\ \\

$\left\|\phi_g^2-(\phi_3^s)^2\right\|_{L^1}$  
&\begin{tabular}{c}
Error \\
Rate \\
\end{tabular} & 
\begin{tabular}{c}
1.054E-1  \\
 0.4470  \\
\end{tabular}& 
\begin{tabular}{c}
7.732E-2 \\
0.4541 \\
\end{tabular} & 
\begin{tabular}{c}
5.644E-2 \\
 0.4579\\
\end{tabular} & 
\begin{tabular}{c}
4.109E-2 \\
0.4649 \\
\end{tabular} & 
\begin{tabular}{c}
2.977E-2  \\
   \\
\end{tabular}  \\ \\

$\left\|\phi_g-\phi_3^s\right\|_{L^2}$  
&\begin{tabular}{c}
Error \\
Rate \\
\end{tabular} & 
\begin{tabular}{c}
2.046E-1 \\
0.2437 \\
\end{tabular} & 
\begin{tabular}{c}
1.728E-1 \\
 0.2461\\
\end{tabular} & 
\begin{tabular}{c}
1.457E-1 \\
0.2455 \\
\end{tabular} & 
\begin{tabular}{c}
1.229E-1  \\
 0.2492  \\
\end{tabular} & 
\begin{tabular}{c}
1.034E-1  \\
   \\
\end{tabular} \\ \\

$\left\|\phi_g-\phi_3^s\right\|_{L^4}$  
&\begin{tabular}{c}
Error \\
Rate \\
\end{tabular}  & 
\begin{tabular}{c}
6.524E-2  \\
 0.3195  \\
\end{tabular} & 
\begin{tabular}{c}
5.228E-2 \\
 0.3241\\
\end{tabular} & 
\begin{tabular}{c}
4.176E-2 \\
 0.3240\\
\end{tabular} & 
\begin{tabular}{c}
3.336E-2 \\
0.3305 \\
\end{tabular} & 
\begin{tabular}{c}
2.653E-2  \\
   \\
\end{tabular} \\ \\

\hline 
\end{tabular}  
\end{center}

Table 7. Convergence rates of the Thomas-Femi approximations
as $\dpm\fl{1}{\kp_3}=\fl{1}{\kp}\to 0$ in 3d ($d=3$). 
\end{table}

\bigskip

From Tables 5-7, we observed numerically 
the convergence rates of
the Thomas-Femi approximations as following: (1) in 1d:
$\left|\mu_g-\mu_1^s\right|\approx O(1/\kp_1^{0.57})$,
$\left\|\phi_g^2-(\phi_1^s)^2\right\|_{L^1}\approx O(1/\kp_1^{0.72})$,
 $\left\|\phi_g-\phi_1^s\right\|_{L^2}\approx O(1/\kp_1^{0.43})$ and 
 $\left\|\phi_g-\phi_1^s\right\|_{L^4}\approx O(1/\kp_1^{0.41})$
as $\fl{1}{\kp_1}\to0$;
(2) in 2d: $\left|\mu_g-\mu_2^s\right|\approx O(1/\kp_2^{0.40})$,
$\left\|\phi_g^2-(\phi_2^s)^2\right\|_{L^1}\approx O(1/\kp_2^{0.58})$,
 $\left\|\phi_g-\phi_2^s\right\|_{L^2}\approx O(1/\kp_2^{0.31})$ and
 $\left\|\phi_g-\phi_2^s\right\|_{L^4}\approx O(1/\kp_2^{0.35})$ 
as $\fl{1}{\kp_2}\to0$; 
and (3) in 3d: $\left|\mu_g-\mu_3^s\right|\approx O(1/\kp_3^{0.31})$,
$\left\|\phi_g^2-(\phi_3^s)^2\right\|_{L^1}\approx O(1/\kp_3^{0.45})$,
 $\left\|\phi_g-\phi_3^s\right\|_{L^2}\approx O(1/\kp_3^{0.24})$ and
 $\left\|\phi_g-\phi_3^s\right\|_{L^4}\approx O(1/\kp_3^{0.32})$
as $\fl{1}{\kp_3}\to0$.

\subsection{Results in 3d with cylindrical symmetry}

The interesting properties of the condensation 
wave function in this case are its root mean square (rms) sizes 
in $r$- and $z$-direction 
$r_{\rm rms}$ and $z_{\rm rms}$, respectively,  defined by
\bea
\label{rrms2d}
&&r_{\rm rms}^2=\langle r^2\rangle=
4\pi\int_0^\ift\int_0^\ift r^2\; \psi^2(r)\;r\;drdz,\\
\label{zrms2d}
&&z_{\rm rms}^2=\langle z^2\rangle=
4\pi\int_0^\ift\int_0^\ift z^2\; \psi^2(r)\;r\;drdz.
\eea

We present computations for two cases:

\bigskip  
 
\noindent {\it Case I. $^{87}$Rb used in JILA with 
$\og_x=\og_y <\og_z$} \cite{Anderson}. The detailed
data are
\beas
&&m=1.44 \tm10^{-25} [kg],\quad \omega_x=\omega_y=20 \pi [1/s], \quad
\omega_z=4 \omega_x [1/s],\\
&&a_0=\sqrt{\fl{\hbar}{\og_x m}}=0.3407 \tm 10^{-5} [m],\ 
|a|=5.1 [nm],\ \kp=4\pi a N/a_0=0.01881 N.
\eeas

\smallskip 
 
\noindent {\it Case II. $^{23}$Na used in MIT (group of Ketterle) with 
$\og_x=\og_y >> \og_z$} \cite{Anglin}.  The detailed
data are
\beas
&&m=3.8 \tm10^{-26} [kg],\quad \omega_x=\omega_y=720 \pi [1/s],\quad  
\omega_z= 7\pi [1/s],\\
&&a_0=\sqrt{\fl{\hbar}{\og_z m}}=1.1209 \tm 10^{-5} [m],\
|a|=2.75 [nm],\ \kp=4\pi a N/a_0=0.003083 N.
\eeas

\bigskip 

In case II, we choose $a_0=\sqrt{\fl{\hbar}{\og_z m}}$ (in stead of 
$a_0=\sqrt{\fl{\hbar}{\og_x m}}$) as $\og_z\ll \og_x$ such that 
the root mean square size is of $O(1)$. The other parameters 
should be adjusted accordingly.

   Figure 4 shows the ground-state condensate wave function
along $r$- and $z$-axis, $\phi_g(r,0)$ and $\phi_g(0,z)$, respectively, 
for different $\kp$ and surface plots of $\phi_g(r,z)$
for $\kp=15408$ and $\kp=188.1$, and  Table 8 lists
$\mu_g$, $\phi_g(0,0)$, $r_{\rm rms}$, $z_{\rm rms}$
 versus $\kp_3=\kp$ for case I. Figure 5 and Table 9
show similar results for case II. 
Furthermore Figure 6 compares the 
numerical ground state solution (i.e. numerical solution of (\ref{mp2d}))
and the Thomas-Fermi approximation (TFA) in (\ref{gssp2d}).
 
\begin{table}[htbp]
\begin{center}
\begin{tabular}{ccccccc}\hline
$a$ &$N$ &$\kp$  &$\phi_g(0,0)$ &$r_{\rm rms}$ &$z_{\rm rms}$  &$\mu_g$ \\
\ &\ &\ &\ &\ &\ &$\ba{cc}
\hbox{Numerical} &\hbox{TFA}\\
\ea$ \\ \hline
-5.1[nm] &250  &-4.705    &0.9926     &0.7468 &0.3268  &$\ba{cc}
\quad 1.9294  &\quad \hbox{NA\ }\ea$ \\
-5.1[nm] &100  &-1.881    &0.6788     &0.9324 &0.3476  &$\ba{cc}
\quad 2.7283  &\quad \hbox{NA\ }\ea$ \\
5.1[nm] &0   &0           &0.602    &1.000  &0.3539 &$\ba{cc}
\quad 3.000  &\quad \hbox{NA\ }\ea$ \\
5.1[nm] &1$\;$000 &18.81  &0.3824     &1.325  &0.3807  &$\ba{cc}
\quad 4.362  &\quad 3.022\ea$\\
5.1[nm] &5$\;$000 &94.05  &0.2477   &1.7742  &0.4214  &$\ba{cc}
\quad 6.680  &\quad 5.752 \ea$\\
5.1[nm] &10$\;$000 &188.1 &0.2023     &2.041 &0.4497  &$\ba{cc}
\quad 8.367  &\quad 7.591\ea$ \\
5.1[nm] &50$\;$000 &940.5 &0.1248     &2.842 & 0.5532  &$\ba{cc}
\quad 14.95  &\quad 14.45\ea$ \\
5.1[nm] &100$\;$000 &1881 &0.1012     &3.276  &0.6174  &$\ba{cc}
\quad 19.47  &\quad 19.06\ea$\\
5.1[nm] &400$\;$000 &7524 &0.0666     &4.341  &0.7881  &$\ba{cc}
\quad 33.47  &\quad 33.19\ea$\\
5.1[nm] &800$\;$000 &15048 &0.0540  &4.992  &8976   &$\ba{cc}
\quad 44.02  &\quad 43.80\ea$\\
 \hline 
\end{tabular}
\end{center}
Table 8. Ground state chemical potential $\mu_g$, maximum value 
of the wave function
$\phi_g(0,0)$ and root mean square sizes $r_{\rm rms}$, $z_{\rm rms}$
versus the interaction coefficient $\kp_3=\kp$ in 3d with cylindrical 
symmetry under case I.
\end{table}

 \begin{table}[htbp]
\begin{center}
\begin{tabular}{ccccccc}\hline
$a$ &$N$ &$\kp$  &$\phi_g(0,0)$ &$r_{\rm rms}$ &$z_{\rm rms}$ &$\mu_g$ \\
\ &\ &\ &\ &\ &\ &$\ba{cc}
\hbox{Numerical} &\hbox{TFA}\\
\ea$ \\ \hline
2.75[nm] &0   &0  &4.3171  &0.0986 &0.7077  &$\ba{cc}
\quad 103.41 &\quad \hbox{NA\ }\ea$ \\
2.75[nm] &1$\;$00 &0.3083 &3.4797   &0.0990  &0.9842  &$\ba{cc}
\quad 104.93 &\quad \hbox{NA\ }\ea$ \\
2.75[nm] &1$\;$000 &3.083  &2.3688   &0.1003 &1.8842 &$\ba{cc}
\quad 111.71 &\quad \hbox{NA\ }\ea$ \\
2.75[nm] &5$\;$000 &15.415  &1.7450   &0.1034 &3.1169 &$\ba{cc}
\quad 127.69 &\quad \hbox{NA\ } \ea$ \\
2.75[nm] &10$\;$000 &30.83  &1.5082  &0.1061  &3.8457 &$\ba{cc}
\quad 141.13  &\quad 86.11 \ea$ \\
2.75[nm] &50$\;$000 &154.15 &1.0202  &0.1179  &6.0508  &$\ba{cc}
\quad 202.04  &\quad 163.9 \ea$\\
2.75[nm] &100$\;$000 &308.3 &0.8416   &0.1267  &7.2226 &$\ba{cc}
\quad 248.1  &\quad 216.3 \ea$ \\
2.75[nm] &500$\;$000 &1541.5 &0.5215   &0.1588  &10.501 &$\ba{cc}
\quad 432.06   &\quad 411.80 \ea$ \\
2.75[nm] &1$\;$000$\;$000 &3083 &0.4226   &0.1784  &12.203 &$\ba{cc}
\quad 559.92   &\quad 543.37 \ea$ \\
2.75[nm] &5$\;$000$\;$000 &15415 &0.2598   &0.2396  &17.070  &$\ba{cc}
\quad 1044.7   &\quad 1034.4 \ea$ \\
 \hline
\end{tabular}
\end{center}
Table 9. Ground state chemical potential $\mu_g$, maximum value 
of the wave function
$\phi_g(0,0)$ and root mean square sizes $r_{\rm rms}$, $z_{\rm rms}$
versus the interaction coefficient $\kp_3=\kp$ in 3d with cylindrical 
symmetry under case II.
\end{table}

\bigskip

 From Figures 4-5, and Tables 8-9, we can see that the chemical 
potential $\mu_g$ and the root mean square sizes $r_{\rm rms}$, 
$z_{\rm rms}$ will increase when 
the interaction coefficient $\kp_3=\kp$ ( i.e. the number of atoms in
the condensate) is increasing. On the other 
hand, the peak of the ground state solution $\phi_g(0,0)$ will decrease.
Figure 6 and Table 8-9 show that  the Thomas-Fermi approximation 
are accurate  for  the chemical potential and 
ground state wave function near the origin only 
when $\kp$ is very big, but gives  
poor approximation when $\kp$ is intermediate
or in the tail of the wave function.

\subsection{Application to compute excited  states of GPE}

 Suppose the eigenfunctions of the nonlinear eigenvalue problem
(\ref{gss})  under the constraint (\ref{normgg}) are
\[\pm\phi_g(\bx), \ \pm\phi_1(\bx),\ \pm\phi_2(\bx),\ \cdots,\]
whose energies satisfy
\[\Eb(\phi_g)<\Eb(\phi_1)<\Eb(\phi_2)<\cdots \;.\]
Then $\phi_j$ is called as the $j$-th excited state solution of 
the GPE (\ref{gpeg}). In fact,
$\phi_g$ and $\phi_j$ ($j=1,2,\cdots$) are critical points 
of the energy functional $\Eb(\phi)$ under the constraint (\ref{normgg}).
In 1d, when $V(x)=\fl{x^2}{2}$ is chosen as the harmonic oscillator 
potential and $\kp_1=0$, the excited states are given \cite{Levine}:
\beas
&&\phi_j(x) = \left(2^j \; j!\right)^{-1/2}\; 
\left(\fl{1}{\pi}\right)^{1/4}e^{-x^2/2}\; H_j(x), \\
&&\mu_j=\mu_0(\phi_j) =E_0(\phi_j)=\left(j+\fl{1}{2}\right),
\qquad j=1,2,\cdots\; ;
\eeas
where $H_j(x)$ is the standard $j$-th Hermite function \cite{Levine}.
Here we show numerically that the algorithm (\ref{cp})
can also be applied to compute any $j$-th excited state of GPE
with $\kp_1>0$ provided that we start with the above
$j$-th excited state as initial data for
$\kp_1>0$ small and use a continuation technique for 
$\kp_1>0$ bigger, i.e. 
use the numerical solution of the $j$-th excited state solution for a
 small $\kp_1$ as initial guess for computing the $j$-th 
excited state of a larger $\kp_1$. When the algorithm (\ref{cp})
is applied to compute $j$-th ($j$ is an odd integer)
excited state in 1d, due to these functions are odd function,
the finite element subspace in Section \ref{s1d}
should be replaced by
\beas
\tilde{V}_g^h&=&\Bigl\{\phi^h(r)\in C([0,R])\ |\  
\phi^h(r)|_{[r_j,r_{j+1}]}\in 
P_1([r_j,r_{j+1}]),\quad 0\le j\le M-1,\\
&&\qquad \phi^h(0)=\phi^h(R)=0\Bigr\}.
\eeas

For simplicity, here we only report numerical results in 1d for the 
first four excited states of GPE for different $\kp_1\ge0$.
  Table 10 shows the energy $\Eb(\phi_j)$ and chemical potential
$\mu_j$ ($j=g, 1,2,3,4$) of the ground state 
and first four excited states of GPE in 1d for different $\kp_1$.
Figure 7 plots the first four excited wave functions $\phi_j(x)$ 
($j=1,2,3,4$)
versus $x$ for different $\kp_1$.

\begin{table}[htbp]
\begin{center}
\begin{tabular}{cccccccc}\hline 
$\kp_1$  &0 &3.1371 &12.5484 &31.371 &62.742 &156.855 &313.71\\ \hline
$\Eb(\phi_g)$ &0.5000 &1.0441 &2.2330 &3.9810 &6.2569 &11.464 &18.171 \\
$\Eb(\phi_1)$ &1.5000 &1.9414 &3.0377 &4.7438 &6.9998 &12.191 &18.891\\
$\Eb(\phi_2)$ &2.5000 &2.8865 &3.9039 &5.5573 &7.7824 &12.944 &19.629\\
$\Eb(\phi_3)$ &3.500 &3.8505 &4.8038 &6.4043 &8.5938 &13.719 &20.383\\
$\Eb(\phi_4)$ &4.500 &4.8245 &5.7252 &7.2752 &9.4276 &14.511 &21.150 \\
\ \\
$\mu_g$ &0.5000 &1.5266 &3.5966 &6.5527 &10.369 &19.070 &30.259 \\
$\mu_1$ &1.5000 &2.3578 &4.3442 &7.2802 &11.089 &19.784 &30.971\\
$\mu_2$ &2.5000 &3.2590 &5.1479 &8.0432 &11.833 &20.512 &31.691\\ 
$\mu_3$ &3.500 &4.1919 &5.9901 &8.8349 &12.597 &21.252 &32.419\\
$\mu_4$ &4.500 &5.1424 &6.8598 &9.6501 &13.381 &22.005 &33.157\\
   \hline
\end{tabular}
\end{center}

Table 10. Energy and chemical potential of the ground state
and first four excited states of GPE in 1d ($d=1$).

\end{table}

\bigskip

From the results in Table 10 and Figure 7, we can see that 
the algorithm (\ref{cp})  can be applied to compute 
the  excited states of GPE (\ref{gpeg}). 
We observed from our numerical results in Table 10
that  for any fixed $\kp_1\ge0$
\[\Eb(\phi_g)<\Eb(\phi_1)<\Eb(\phi_2)<\cdots,
\qquad \mu_g<\mu_1< \mu_2<\cdots\;. \]
This implies that the eigenvalue of (\ref{gss}), $\mu_g$,  
corresponding to the minimizer of the energy functional 
$\Eb(\phi)$, $\phi_g$,
is the minimum chemical potential among 
all the eigenvalues of (\ref{gss}).  
Furthermore, we have
\[\lim_{\kp_1\to +\ift}\ \fl{\Eb(\phi_j)}{\Eb(\phi_g)}=1, \qquad
 \lim_{\kp_1\to +\ift}\ \fl{\mu_j}{\mu_g}=1, 
\qquad j=1,2,3,4.\]
A rigorous mathematical justification of these numerical 
observations is under further study.

\section{Conclusions}\label{sc}  
\setcounter{equation}{0}

  Ground state solution of time-independent Gross-Pitaevskii equation 
of Bose-Einstein condensation 
at zero or very low temperature is computed by directly
 minimizing the energy functional under a constraint
through the finite element discretization.  
We begin with the 3d Gross-Pitaevskii equation,
scale it to obtain a three-parameter model, show how to reduce
it to 2d and 1d GPEs.
The ground state solution is formulated via minimizing the energy functional
under a constraint. The finite element approximation for 1d, 2d with radial
symmetry and 3d with spherical symmetry and
cylindrical symmetry  are presented in detail and 
approximate ground state solutions, which are used as
initial guess in our practical numerical computation,  are provided
in two extreme
regimes: very weak interactions and strong repulsive interactions.
Numerical results are reported in 1d, 2d with radial symmetry 
and 3d with spherical symmetry and cylindrical symmetry for 
condensation with repulsive/attractive interparticle interactions
and atoms in it ranging up to millions 
 to demonstrate the novel numerical method. 
Our numerical results show that the Thomas-Fermi approximation
are accurate  for  the chemical potential and 
ground state wave function near the origin only 
when $\kp$ is very big, but gives  
poor approximation when $\kp$ is intermediate
or in the tail of the wave function.
Furthermore extension of our method to compute the
excited states of GPE is also presented.
 In the future we plan to study physically more complex systems 
based on this ground state solution solver.

\bigskip      
 
\begin{center} 
{\large \bf Acknowledgment}  
\end{center} 
 
W.B. acknowledges support  by the National University of Singapore
 grant No. R-151-000-027-112 
and helpful discussion with 
Peter A. Markowich.
W. T. acknowledges support  by the National University of Singapore,
 hospitality by Guowei Wei and Yingfei Yi for his extended
visit at Department of Computational Science, National University
of Singapore, and support by the National Natural Science
Foundation of China grant No. 19901004-1. 
The authors thank the referees for their very helpful suggestions
and comments to improve the paper.

\newpage  
  
 \begin{figure}[htb]  
\centerline{a). \psfig{figure=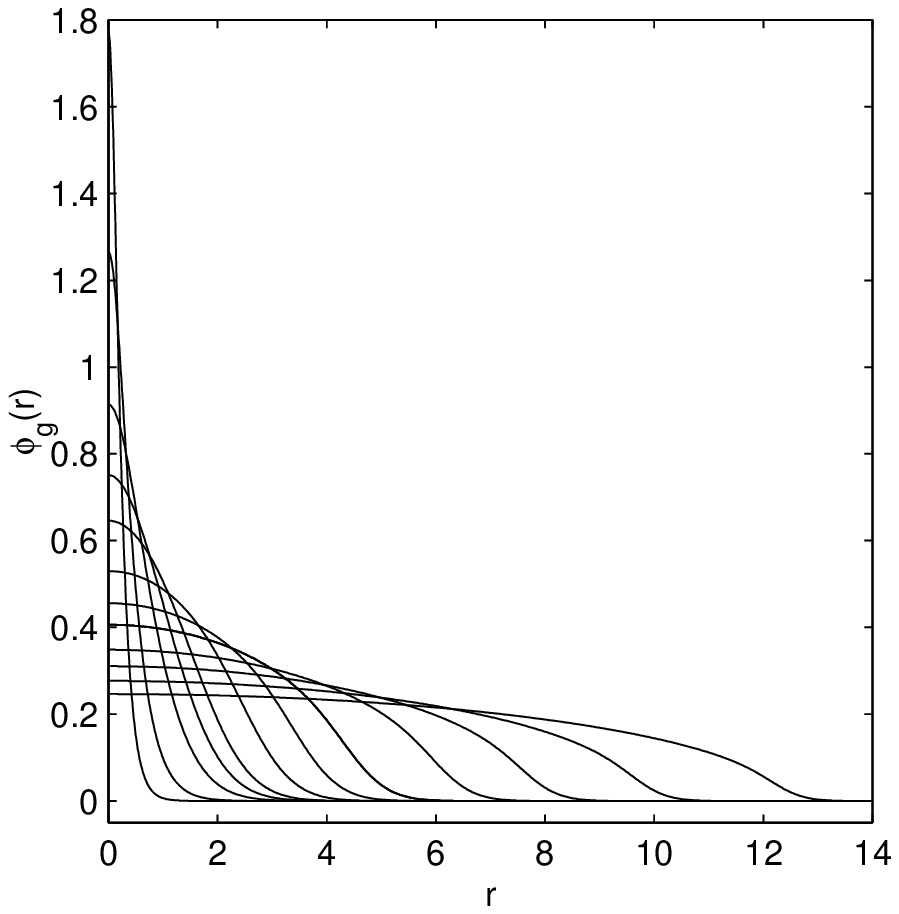,height=8cm,width=7cm,angle=0}
\qquad b). \psfig{figure=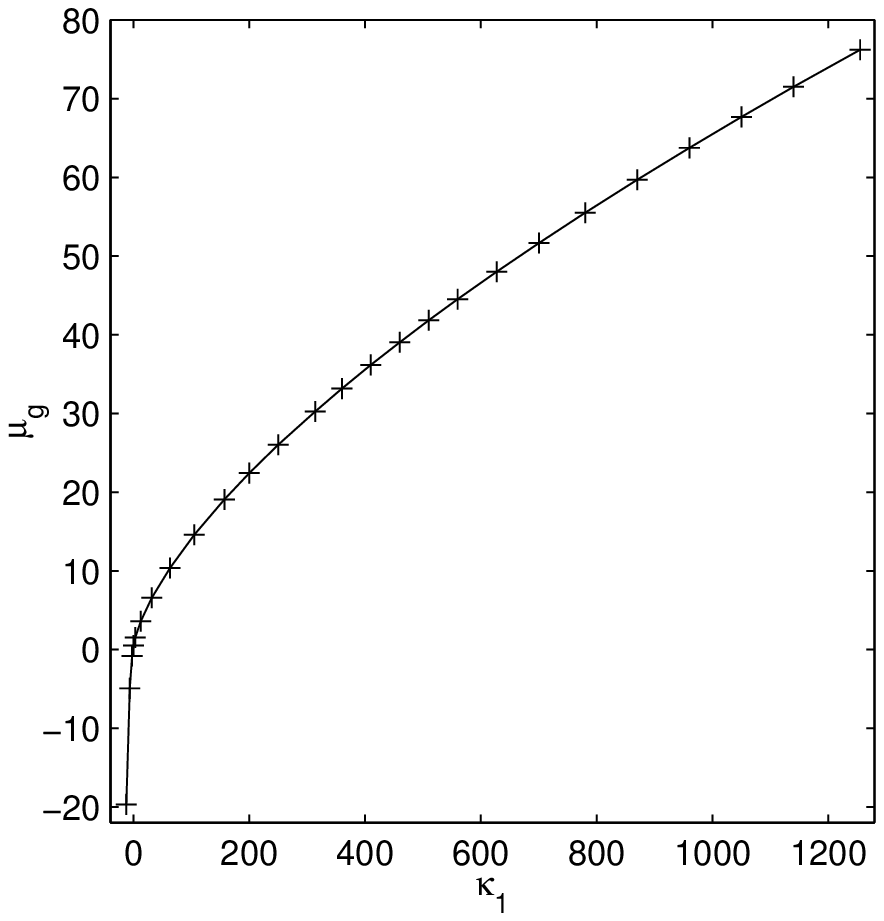,height=8cm,width=7cm,angle=0}}   
 Figure 1: Ground-state condensate solution in 1d ($d=1$).
a). Wave function $\phi_g(r)$ versus $r$
for $\kp_1=-12.5484$, $-6.2742$, $-2.5097$, $0$, $3.1371$, $12.5484$,
$31.371$, $62.742$,  $156.855$, $313.71$, $627.42$, $1254.8$ 
(in order of increasing width).  b). Chemical potential 
$\mu_g$ versus $\kp_d$. 

\bigskip
\bigskip
\bigskip

\centerline{a). \psfig{figure=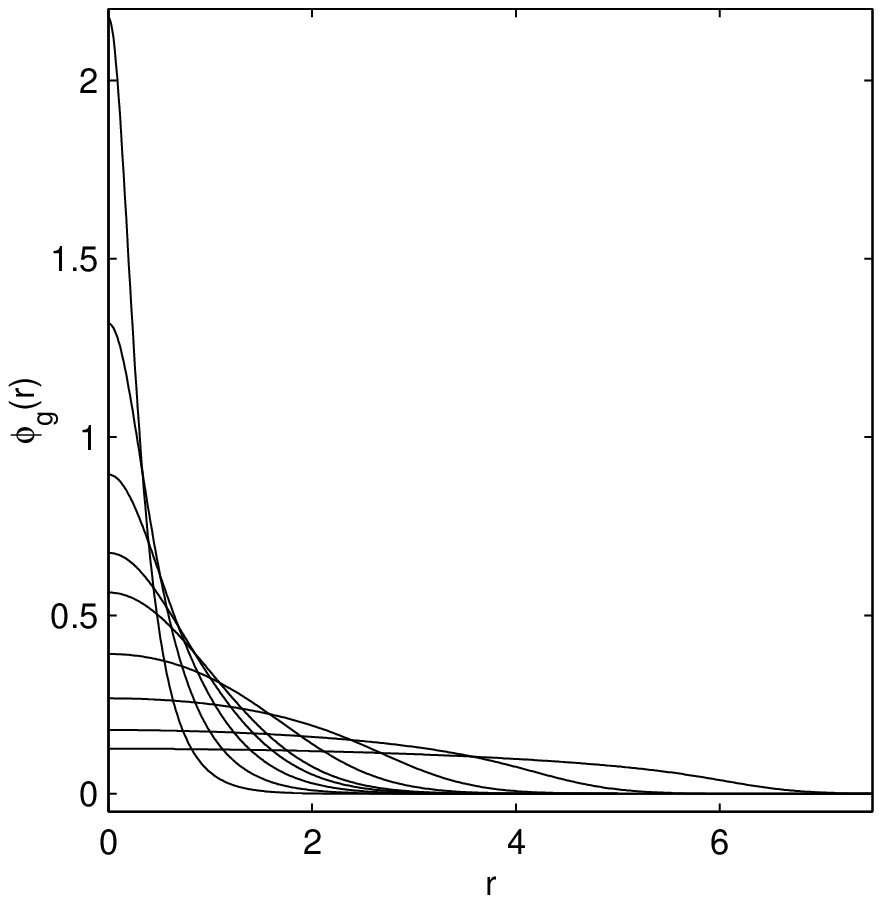,height=8cm,width=7cm,angle=0}
\qquad b). \psfig{figure=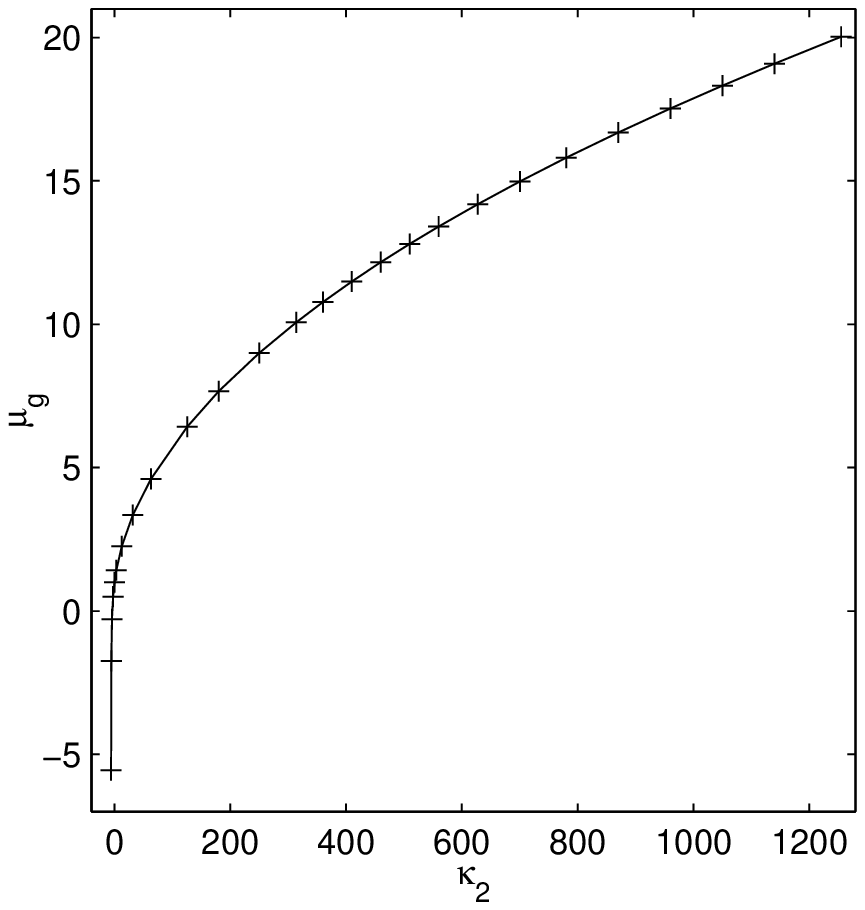,height=8cm,width=7cm,angle=0}}   
  Figure 2: Ground-state condensate solution in 2d with radial symmetry.
a). Wave function $\phi_g(r)$ versus $r$
 for $\kp_2=-5.8$, $-5.5$, $-4.5$, $-2.5097$, $0$, $12.5484$,
$62.742$, $313.71$, $1254.8$ (in order of increasing width). 
b). Chemical potential $\mu_g$ versus $\kp_d$

\end{figure} 

\clearpage

\newpage  
  
 \begin{figure}[htb]  
 \centerline{a). \psfig{figure=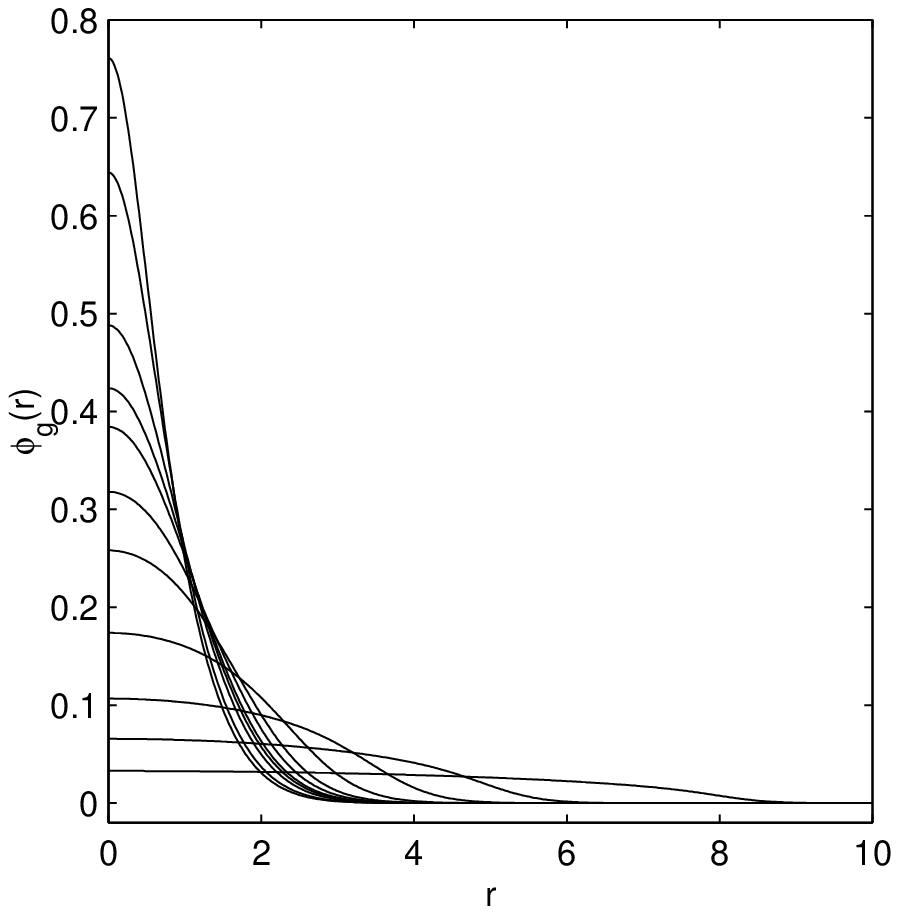,height=8cm,width=7cm,angle=0}
\qquad b). \psfig{figure=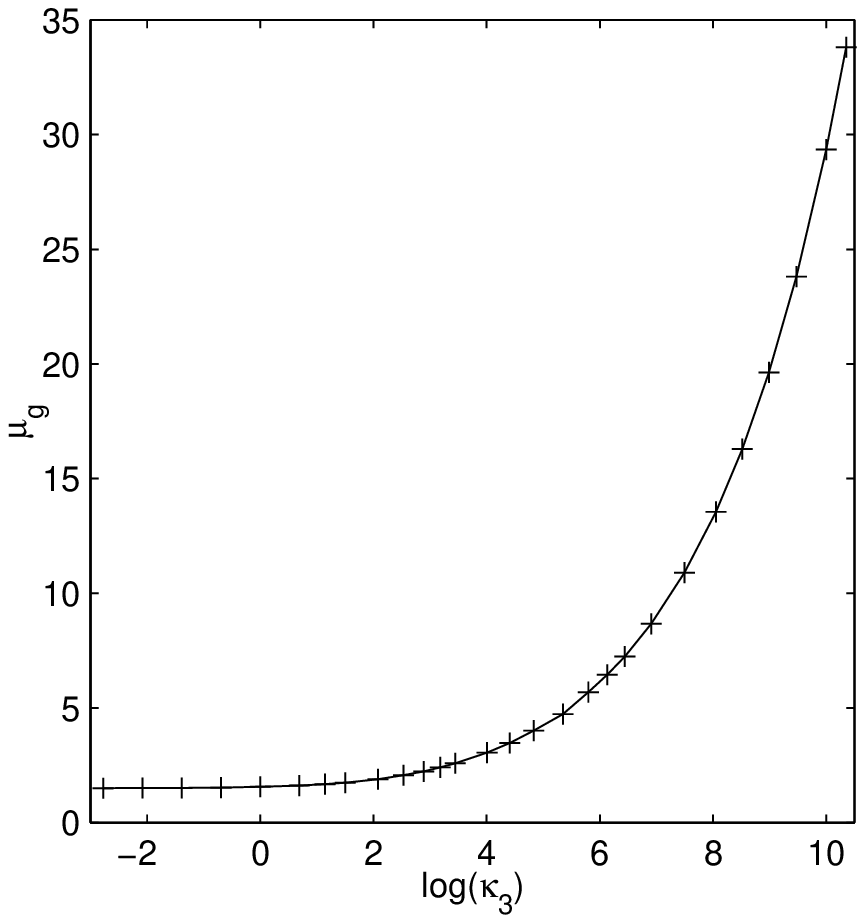,height=8cm,width=7cm,angle=0}}   
  Figure 3: Ground-state condensate solution in 3d with spherical symmetry. 
a). Wave function $\phi_g(r)$ versus $r$
for $\kp_3=-7$, $-6.2472$, $-3.1371$, $0$, $3.1371$,  $12.5484$,
$31.371$, $125.484$,  $627.42$, $3137.1$, $31371$ 
(in order of increasing width). b). Chemical potential $\mu_g$ versus $\kp_d$.
 \end{figure}

 \begin{figure}[htb]  
\centerline{a).\psfig{figure=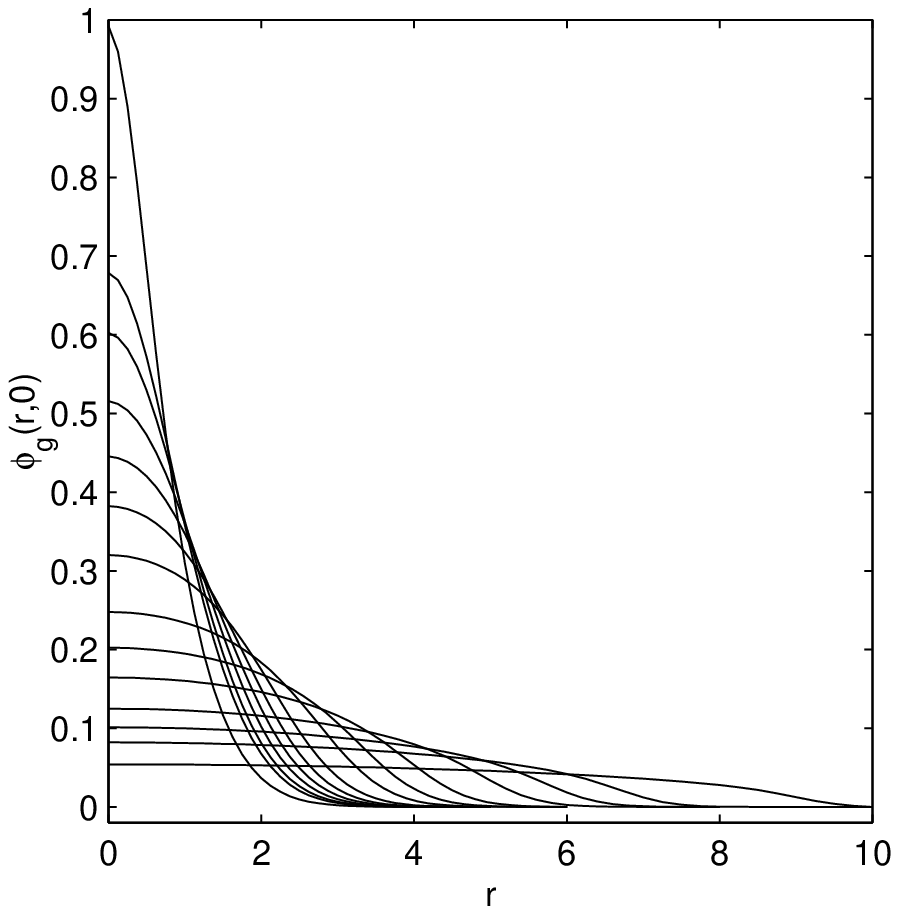,height=8cm,width=7cm,angle=0}
\qquad b).\psfig{figure=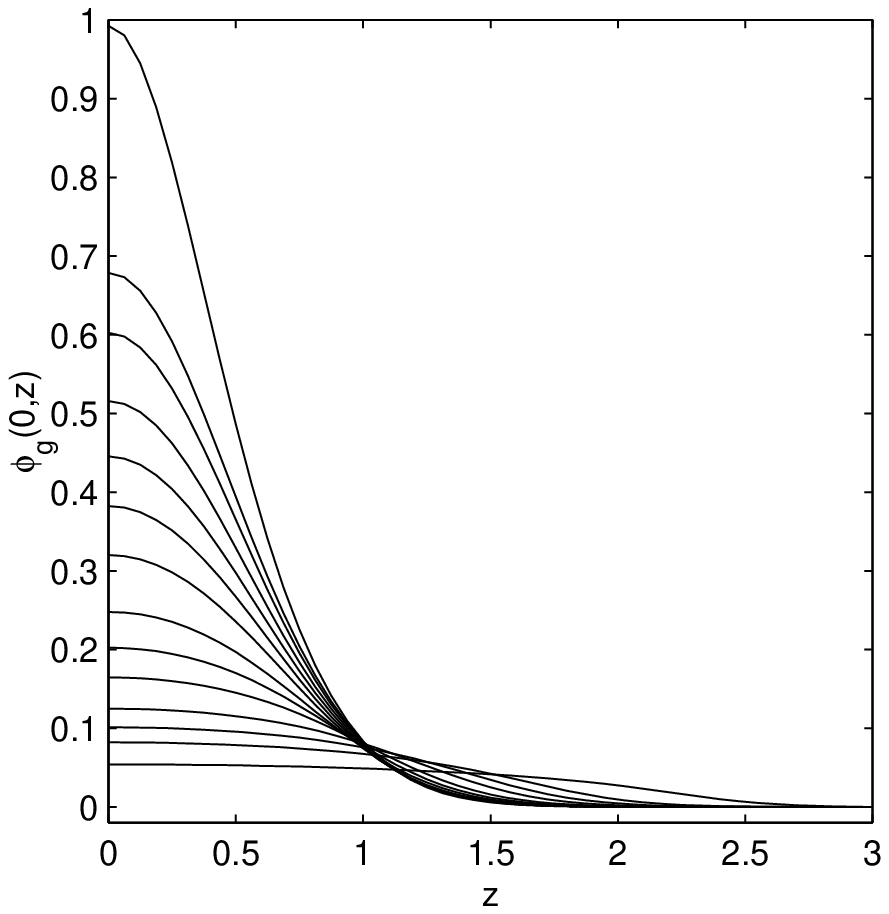,height=8cm,width=7cm,angle=0}}   

\vspace{1cm} 
\centerline{c).\psfig{figure=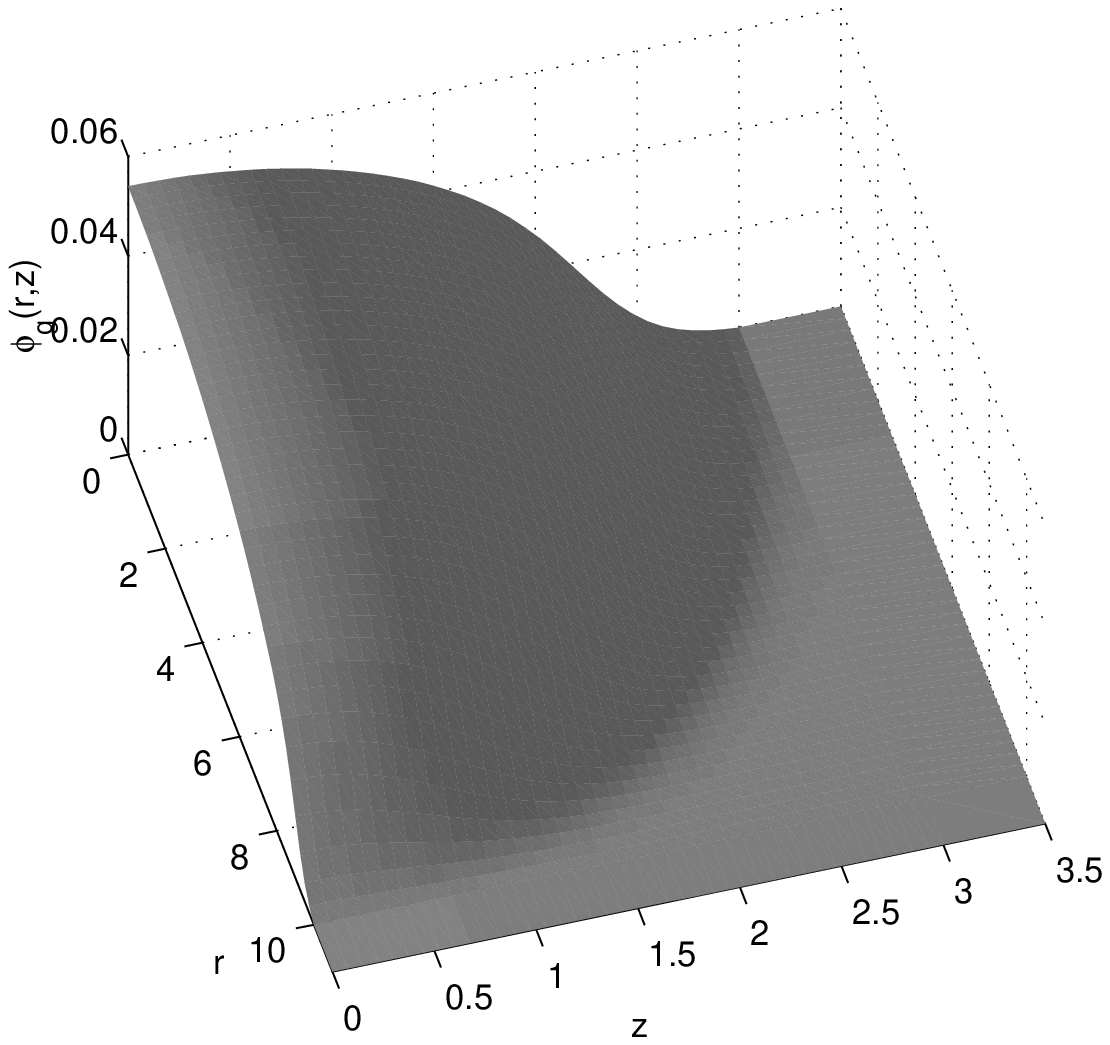,height=8cm,width=7cm,angle=0}
\qquad d).\psfig{figure=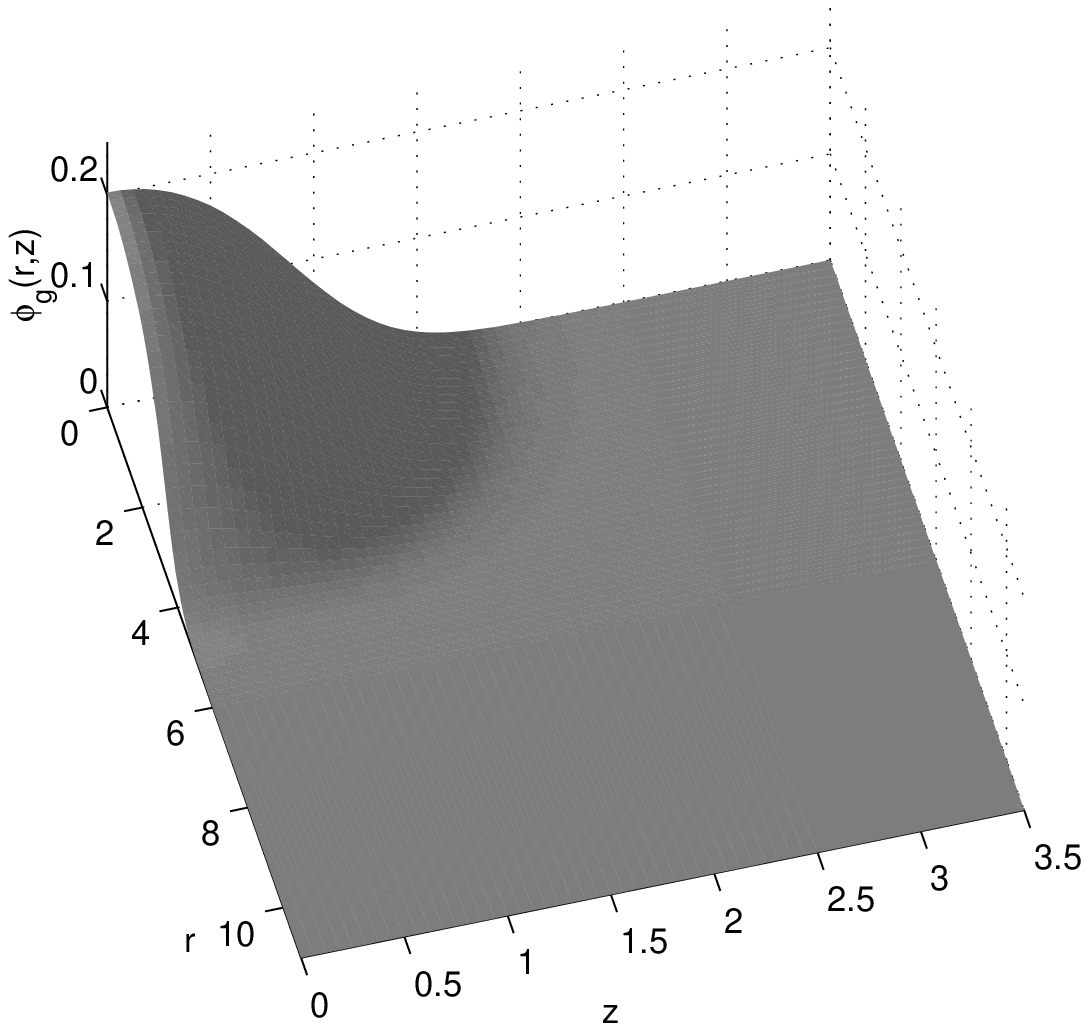,height=8cm,width=7cm,angle=0} }   

Figure 4: Ground-state solution in 3d with cylindrical symmetry under case I.
Condensate wave function  on two lines  for 
 $\kp_3=-4.705$, $-1.881$, $0$, $3.762$, $9.405$,
$18.81$, $37.62$,  $94.05$, $188.1$, $376.2$, $940.5$, $1881$, $ 3762$ and
$15048$ (in order of increasing width):  a). On the line $z=0$ 
($\phi_g(r,0)$);
b).  On the line $r=0$ ($\phi_g(0,z)$).
Surface Plots of the condensate wave function 
$\phi_g(r,z)$: 
  c). $\kp_3=15048$ ($N=800\;000$);
d). $\kp_3=188.1$ ($N=10\;000$). 
\end{figure}

 \begin{figure}[htb]  
\centerline{a).\psfig{figure=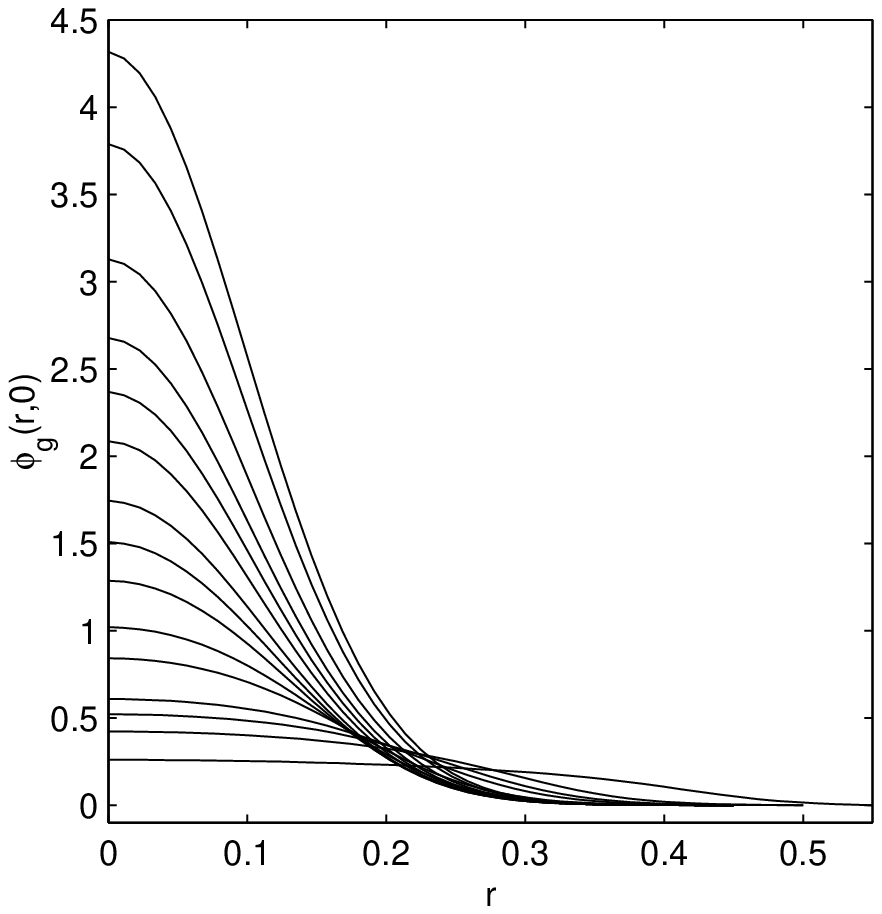,height=8cm,width=7cm,angle=0}
\qquad b).\psfig{figure=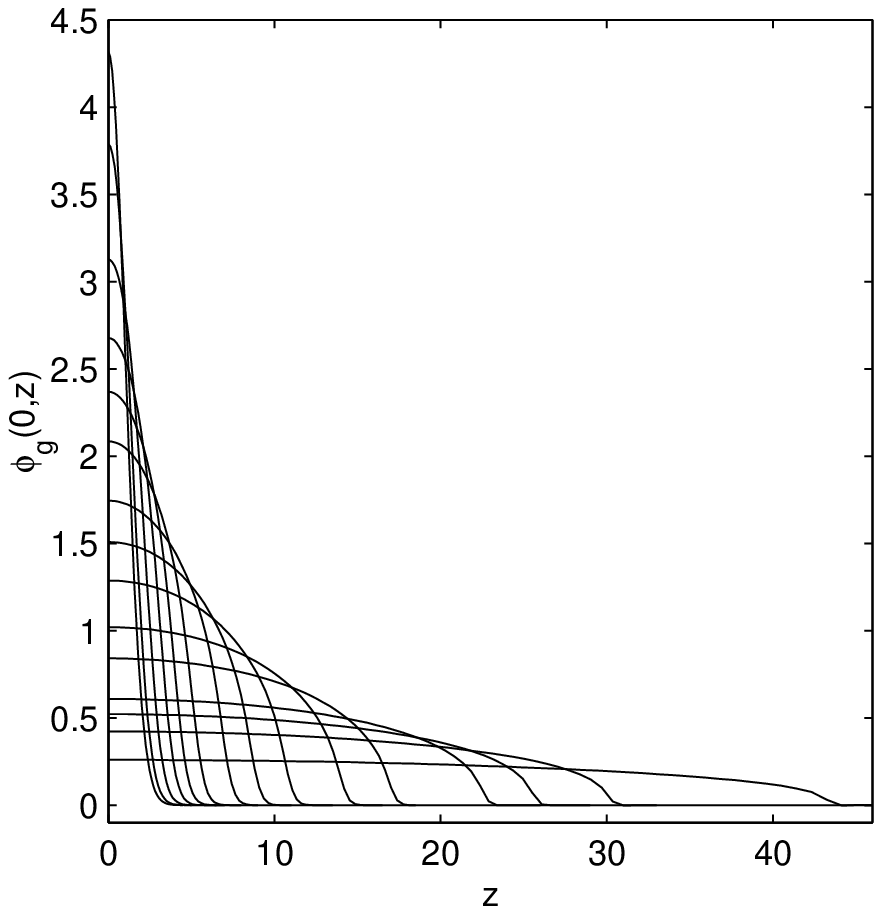,height=8cm,width=7cm,angle=0} }   

\vspace{1cm} 
\centerline{c).\psfig{figure=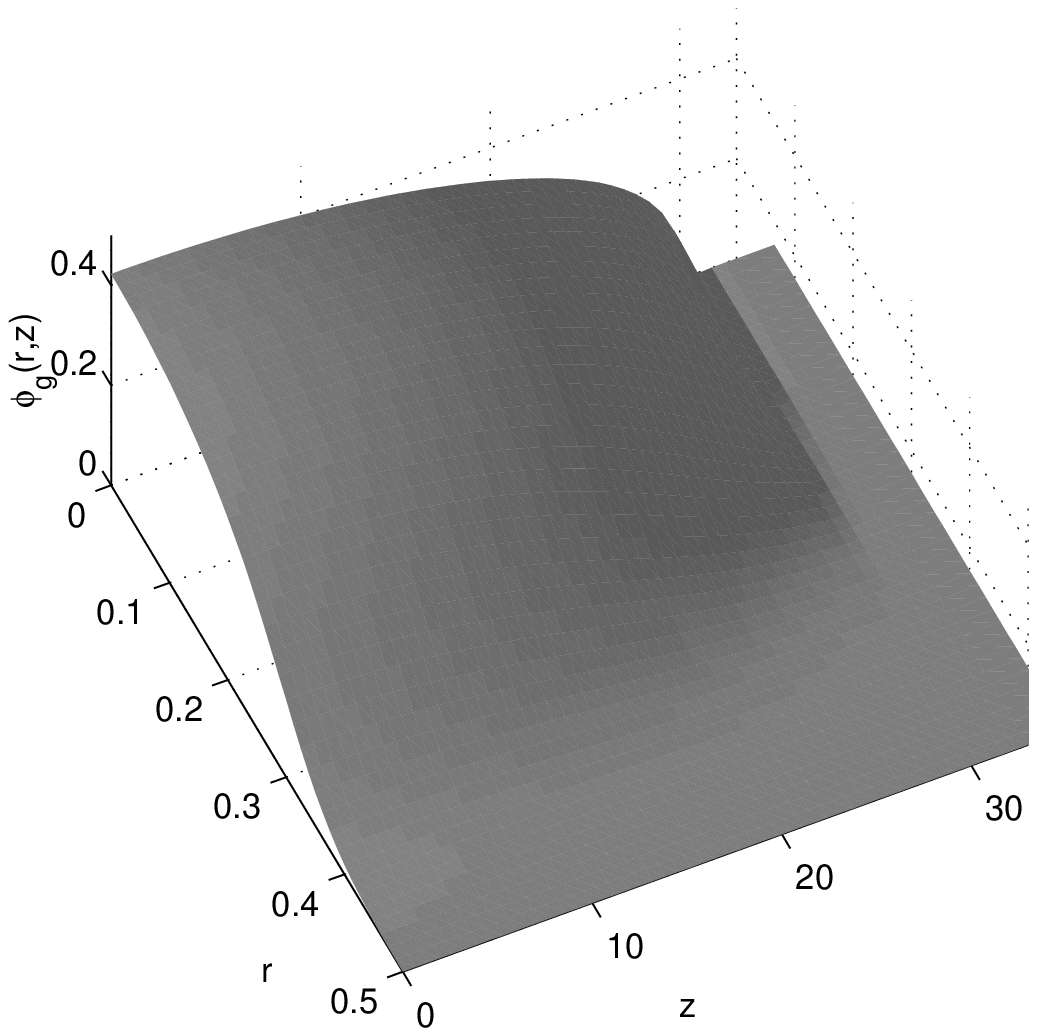,height=8cm,width=7cm,angle=0}
\qquad d).\psfig{figure=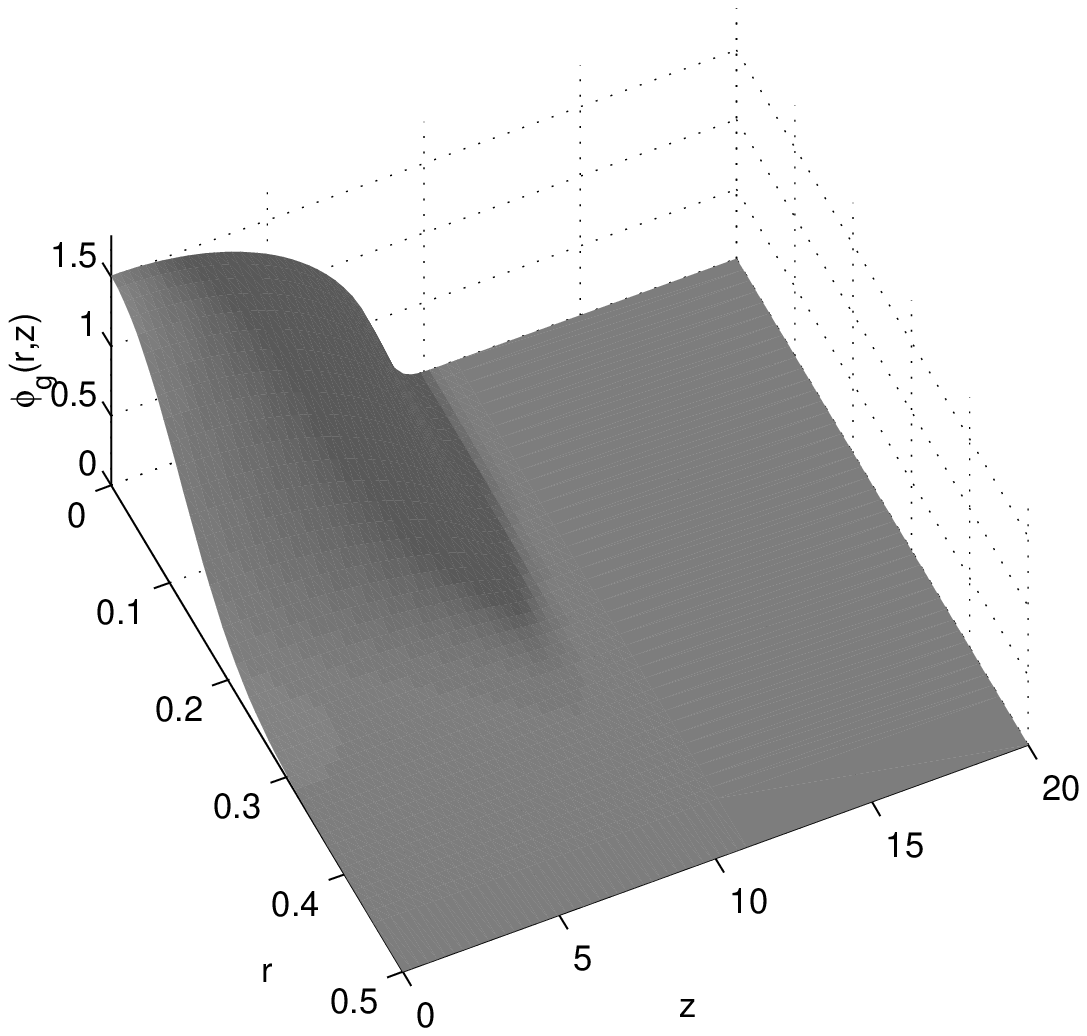,height=8cm,width=7cm,angle=0}}

Figure 5: Ground-state solution in 3d with cylindrical symmetry under case II.
Condensate wave function  on two lines  for $\kp_3=0$, $0.15415$, $0.6166$,
$1.5415$, $3.083$,  $6.166$, $15.415$, $30.83$, $61.66$, $154.15$, 
$308.3$, $924.9$, $1541$, $3083$ and $15415$
  (in order of increasing width):  a). On the line $z=0$ ($\phi_g(r,0)$);
b).  On the line $r=0$ ($\phi_g(0,z)$).
Surface Plots of the condensate wave function 
$\phi_g(r,z)$: 
  c). $\kp_3=3083$ ($N=1\;000\;000$);
d). $\kp_3=30.83$ ($N=10\;000$). 
 
\end{figure}

\clearpage

 \begin{figure}[htb]  
\centerline{a).\psfig{figure=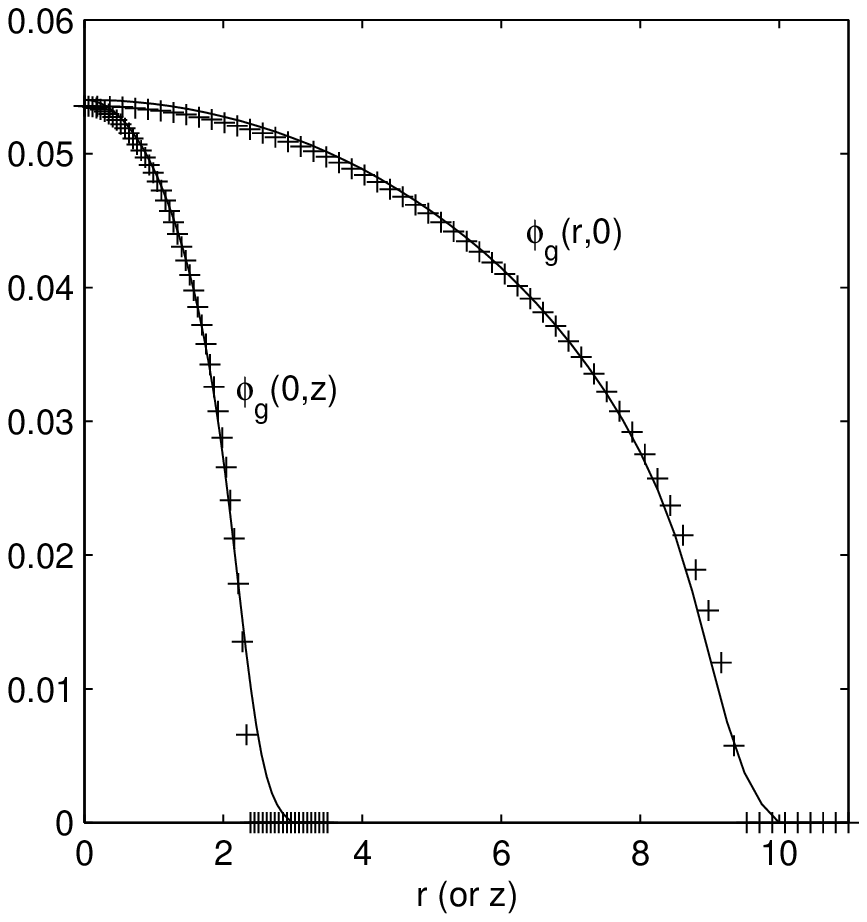,height=8cm,width=7cm,angle=0}
\qquad b).\psfig{figure=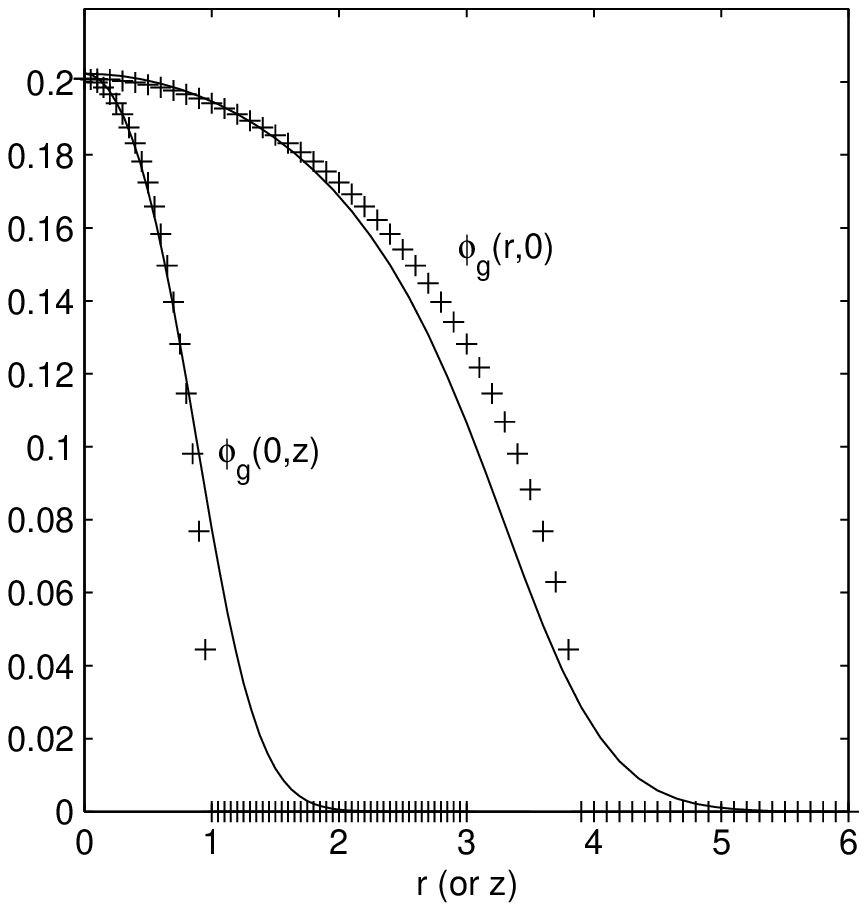,height=8cm,width=7cm,angle=0}}   

\vspace{1cm} 
\centerline{c).\psfig{figure=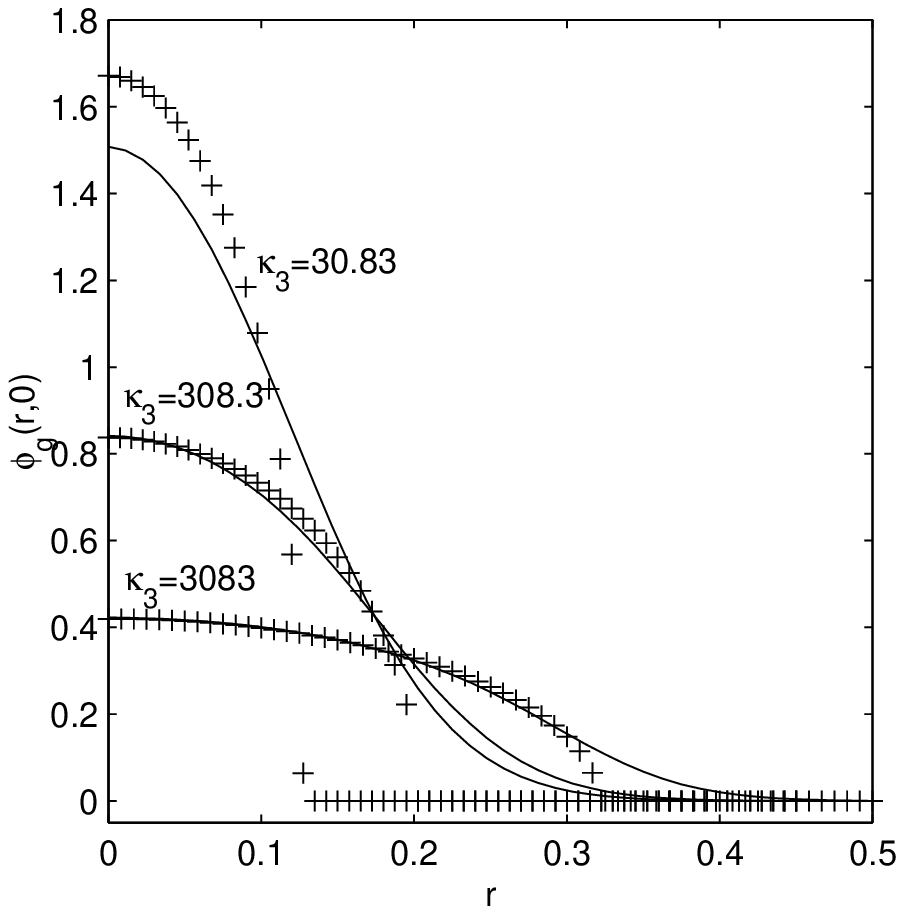,height=8cm,width=7cm,angle=0}
\qquad d).\psfig{figure=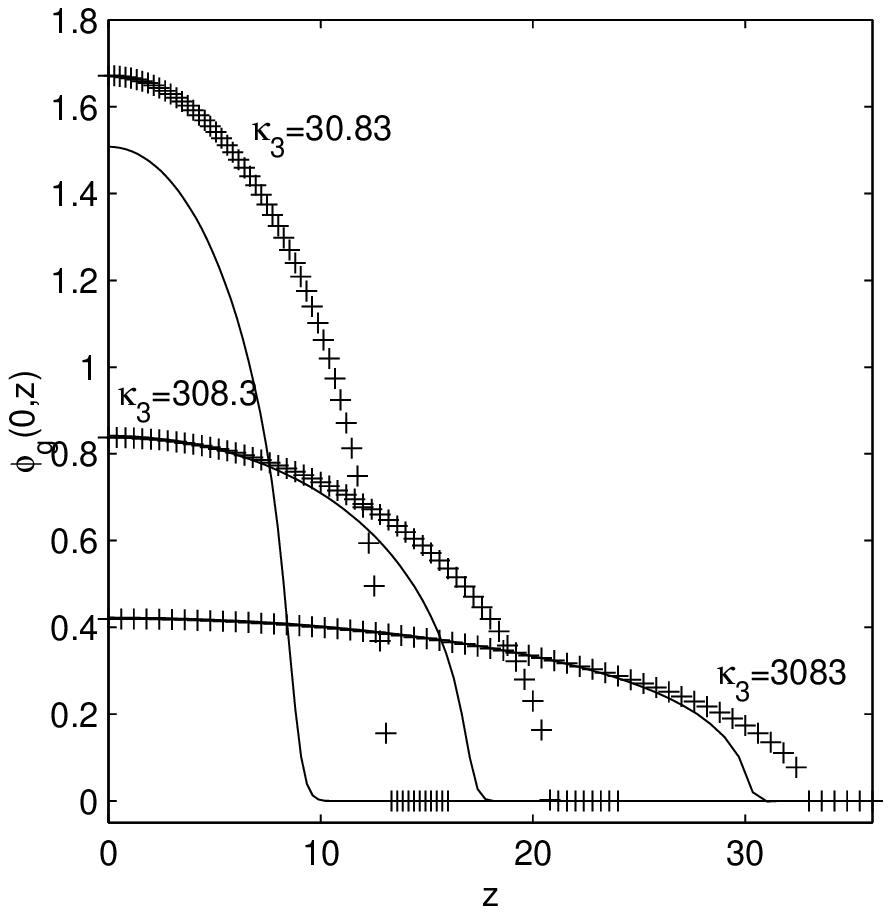,height=8cm,width=7cm,angle=0}}   

\vspace{1cm} 

 Figure 6: Comparison between the numerical ground state solution and 
the Thomas-Femi approximation in 3d with cylindrical symmetry. 
`---': Numerical solution of (\ref{mp2d}), `+ + +': 
Thomas-Fermi approximation (\ref{gssp2d}).
For case I: a). $\kp_3=15048$ ($N=800\;000$);
b). $\kp_3=188.1$ ($N=10\;000$).
For case II:  c). At the line $z=0$, i.e. $\phi_g(r,0)$;
d). At the line $r=0$, i.e. $\phi(0,z)$.  
\end{figure}

\newpage
\begin{figure}[htb]  
\centerline{a).\psfig{figure=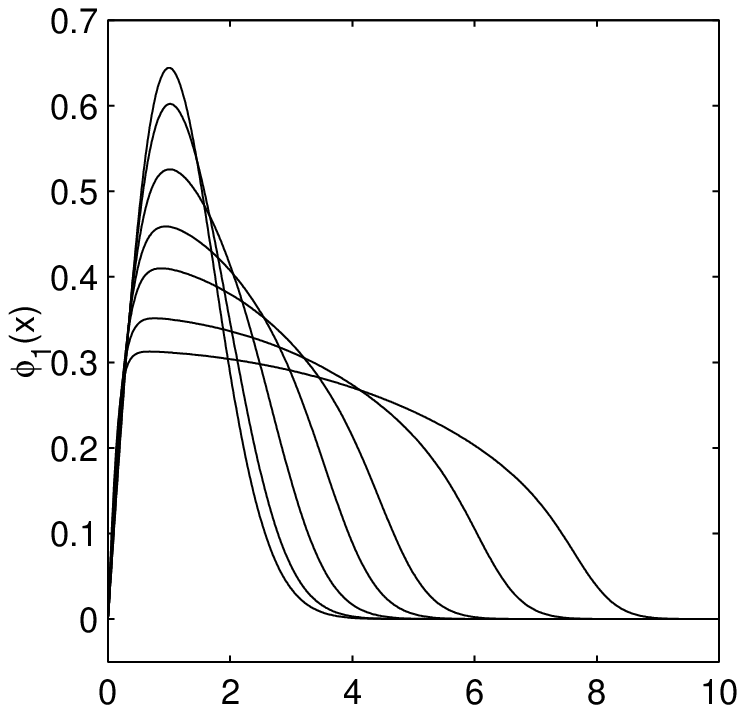,height=8cm,width=7cm,angle=0}
\qquad b).\psfig{figure=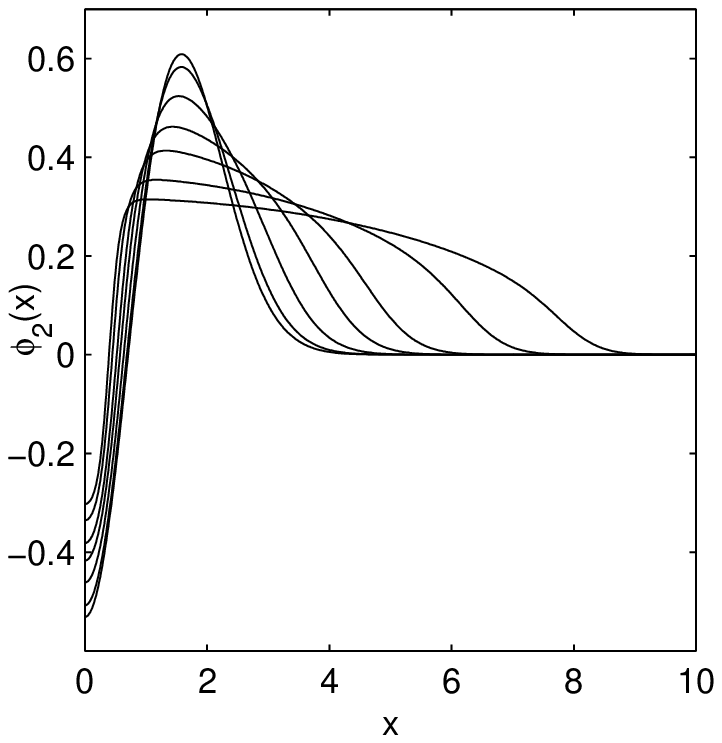,height=8cm,width=7cm,angle=0}}   

\vspace{1cm} 
\centerline{c).\psfig{figure=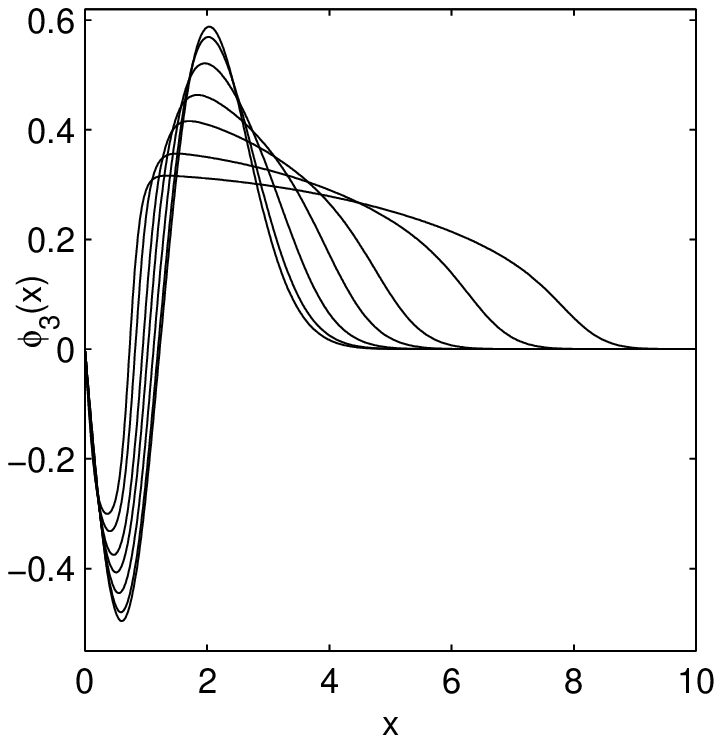,height=8cm,width=7cm,angle=0}
\qquad d).\psfig{figure=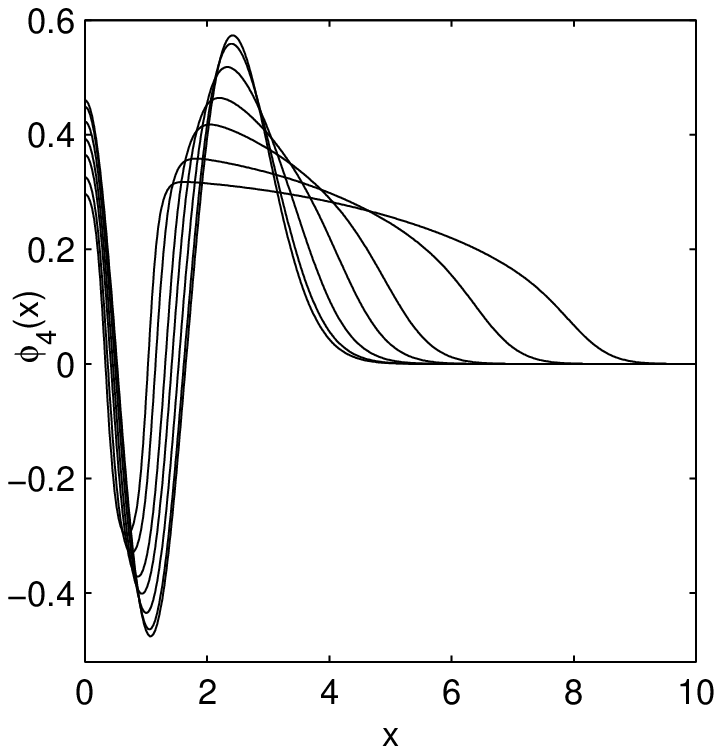,height=8cm,width=7cm,angle=0}}   

\vspace{1cm} 

 Figure 7: Excited states of GPE with harmonic oscillator potential
in 1d for  $\kp_1=0$, $3.1371$, $12.5484$,
$31.371$, $62.742$,  $156.855$, $313.71$
(in order of increasing width).
a). First excited state $\phi_1(x)$ (odd function);
b). Second excited state $\phi_2(x)$ (even function);
c). Third excited state $\phi_3(x)$ (odd function);
d). Firth excited state $\phi_4(x)$ (even function).

\end{figure}

\end{document}